\documentclass[11pt,eqno]{article}
\usepackage{amsmath}
\usepackage{amsthm}
\usepackage{amssymb}
\usepackage{amsfonts}
\usepackage{epsf}
\usepackage{graphicx}
\usepackage{epstopdf}
\usepackage{hyperref}
\usepackage{color}

\def\Im {\mathop{\rm Im}\nolimits}
\def\arg {\mathop{\rm arg}\nolimits}
\def\Re {\mathop{\rm Re}\nolimits}
\def\Ai {{\rm Ai}}

\newtheorem{pro}{PROPOSITION}
\newtheorem{cor}{COROLLARY}
\newtheorem{lem}{LEMMA}
\newtheorem{thm}{THEOREM}

\newtheorem{rem}{REMARK}

\numberwithin{equation}{section}

\title{{Critical edge behavior and the Bessel to Airy transition in the singularly perturbed Laguerre   unitary  ensemble}}

\author{Shuai-Xia Xu$^a$, Dan Dai$^b$ and Yu-Qiu Zhao$^c$\footnote{Corresponding author (Yu-Qiu Zhao).
 {\it{E-mail
address:}} {stszyq@mail.sysu.edu.cn} }}
  \date{
 {\it{$^a$Institut Franco-Chinois de l'Energie Nucl\'{e}aire, Sun Yat-sen University, GuangZhou
510275,  China}}\\
{\it{$^b$Department of Mathematics, City University of Hong Kong, Tat Chee Avenue, Kowloon, Hong Kong}}\\
 {\it{$^c$Department of Mathematics, Sun Yat-sen University, GuangZhou
510275, China}}
}

\begin{document}

\maketitle

\noindent \hrule width 6.27in\vskip .3cm

\noindent {\bf{Abstract }}

In this paper, we study the singularly perturbed Laguerre unitary ensemble
$$
\frac{1}{Z_n} (\det M)^\alpha e^{- \textrm{tr}\, V_t(M)}dM, \qquad  \alpha >0,
$$ with $V_t(x) = x + t/x$, $x\in (0,+\infty)$ and $t>0$. Due to the effect of $t/x$ for varying $t$, the eigenvalue correlation kernel has a new limit instead of the usual Bessel kernel at the hard edge 0. This limiting kernel involves $\psi$-functions associated with a special solution to a new third-order nonlinear differential equation, which is then shown equivalent to a particular Painlev\'e III equation. The  transition of this limiting kernel to the Bessel and Airy kernels is also studied when the parameter $t$ changes in a finite interval $(0, d]$. Our approach is based on  Deift-Zhou nonlinear steepest descent method for Riemann-Hilbert problems.
  \vskip .5cm
 \noindent {\it{2010 Mathematics subject classification:}} 33E17; 34M55; 41A60

\vspace{.2in} \noindent {\it {Keywords: }}Random matrices; perturbed Laguerre unitary ensemble;
Riemann-Hilbert approach; uniform asymptotic approximation; Painlev\'{e} III equation
\vskip .3cm

\noindent \hrule width 6.27in\vskip 1.3cm

\newpage

\section{Introduction and statement of results} \setcounter{section} {1}

For $n \in \mathbb{N}$, $\alpha >0$  and $t>0$, we consider the following unitary random matrix ensemble
\begin{equation}\label{vt-prob-measure}
\frac{1}{Z_n} (\det M)^\alpha e^{- \textrm{tr}\, V_t(M)}dM,\ \ \
dM=\prod_{i=1}^{n}dM_{ii}\prod_{i=1}^{n-1}\prod_{j=i+1}^{n}d\Re
M_{ij}d\Im M_{ij}
\end{equation}
on the space of $n\times n$ positive definite  Hermitian  matrices $M=(M_{ij})_{n\times n}$. Here
\begin{equation} \label{partition-function}
  Z_n=\int (\det M)^\alpha e^{- \textrm{tr} V_t(M)}dM
\end{equation}
is the normalization constant,  and
\begin{equation} \label{vt-def}
  V_t(x):= x + \frac{t}{x}, \qquad x \in (0,\infty), \quad t >0.
\end{equation}
When $t=0$, we have $V_0(x) = x$,  and \eqref{vt-prob-measure} is reduced to the well-known Laguerre unitary ensemble (LUE); see, e.g.,  Forrester \cite[Chap. 3]{Forrester-book}. In this paper, by introducing the extra term $t/x$ in \eqref{vt-def}, we call \eqref{vt-prob-measure} the perturbed Laguerre unitary ensemble (pLUE). We note that this pLUE has recently been  considered by Chen and Its \cite{ci}, where a relation with the
Painlev\'e III (PIII, for short) function was discovered.

It is well-known, see e.g. \cite{deift,Forrester-book,mehta}, that the eigenvalue correlation kernel for the ensemble \eqref{vt-prob-measure} has the following form
\begin{equation}\label{kernel-formula}
K_n(x,y;t)= x^{\frac{\alpha}{2}} y^{\frac{\alpha}{2}} e^{- \frac{V_t(x)+
V_t(y)} 2}\sum_{k=0}^{n-1}p_k(x)p_k(y),
\end{equation}
where $p_k(x)$ denotes the $k$-th degree orthonormal polynomial with
respect to the weight
\begin{equation}\label{weight-of-the-paper}
w(x)=w(x;t)=x^{\alpha}e^{-V_t(x)},~~~x\in (0, \infty),~~t>0,~~\alpha>0.
\end{equation}
Using the famous Christoffel-Darboux formula, \eqref{kernel-formula} can be put into the following closed form
\begin{equation} \label{kernel-CD-formula}
  K_n(x,y;t)=\gamma^2_{n-1} \sqrt{w(x) w(y)}
\frac {\pi_n(x)\pi_{n-1}(y)-\pi_{n-1}(x)\pi_n(y)}{x-y},
\end{equation}
where $\gamma_k$ is the leading coefficient of $p_k(x)$, and $\pi_k(x)$ is the monic polynomial such that $p_k(x)=\gamma_k \pi_k(x)$.

In the study of random matrices, there is a lot of interest in the limit of the correlation kernel $K_n(x,y)$ when the matrix size $n$ tends to infinity. For the LUE case ($t=0$), the limiting mean eigenvalue density is
\begin{equation} \label{MP-measure}
  \psi_V(x) = \lim_{n\to \infty} 4 K_n(4nx,4nx;0) = \frac{2}{\pi} \sqrt{\frac{1-x}{x}}, \qquad x\in (0,1);
\end{equation}
see e.g. \cite[p.106]{Forrester-book}. Note that the above density is independent of $\alpha$, and this is a typical example of the  Mar\v{c}enko-Pastur law; see \cite{Mar:Pas}. Moreover, it is well-known that the limiting behavior of $K_n$
is given by the sine kernel
\begin{equation} \label{sine-kernel}
   \mathbb{S}(x,y):=\frac{\sin \pi (x-y)}{x-y}
\end{equation}
in the bulk of the spectrum \cite{Fox,Nagao}, by the Airy kernel
\begin{equation} \label{airy-kernel}
   \mathbb{ A}(x,y):= \frac{\Ai(x) \Ai'(y) - \Ai(y) \Ai'(x)}{x-y}
\end{equation}
at the soft edge of the spectrum \cite{Forrester1993,Tra:Wid1}, and by the Bessel kernel
\begin{equation} \label{bessel-kernel}
  \mathbb{ J}_{\alpha}(x,y):=\frac{J_\alpha(\sqrt{x}) \sqrt{y} J'_\alpha(\sqrt{y}) - J_\alpha(\sqrt{y}) \sqrt{x} J'_\alpha(\sqrt{x})}{2(x-y)}
\end{equation}
at the hard edge of the spectrum \cite{Forrester1993,Tra:Wid2}. In general, when $V_0(x)$ is a polynomial instead of the simplest monomial $x$ in \eqref{vt-def}, Vanlessen \cite[Thm. 2.7]{vanlessen} proved that the above limiting kernels hold as well. Of course, one may consider an even more general case by assuming that $V_0(x)$ is a real analytic function and satisfies
\begin{equation*}
  \lim_{x \to +\infty} \frac{V_0(x)}{\ln(x^2+1)} = +\infty,
\end{equation*}
and the above limiting kernels are also expected.  Such a phenomenon  is called universality in random matrix theory.
To prove these universality results, the Deift-Zhou nonlinear steepest descent method is a very powerful tool; for example, see \cite{deift,dkmv1,kv}.
It is also worth   pointing  out that, some kernels involving Painlev\'e functions have  appeared  in certain critical situations, see e.g. \cite{bi,ck2008,CKV,ik1,xzz2011}.

In the present  paper, we will focus on the case when  $V_t(x)$ is not real analytic. To be precise,  we will work on  one of the simplest non-analytic cases, in which  $V_t(x)$ possess a simple pole at the hard edge $0$; cf. \eqref{vt-def}. The exponent  $t/x$   induces an infinitely strong zero of the weight $w(x;t)$ at the origin; cf. \eqref{weight-of-the-paper}. Therefore, it is natural to expect that the distribution  of eigenvalues near $0$ will change dramatically   due to the perturbation   $e^{-t/x}$, and the limiting kernel at the hard edge may no longer be the Bessel kernel $\mathbb{ J}_{\alpha}$ in \eqref{bessel-kernel}. Indeed, we will show that the limiting kernel is related to a  \emph{third-order} nonlinear differential equation. The third-order nonlinear differential equation is integrable and its Lax pair and the corresponding Riemann-Hilbert problem (RH problem or RHP, for short)  will be provided in Section \ref{sec:2} below. With the initial values adapted,
later in Section \ref{sec:2.2}, the third-order equation is shown equivalent to a particular PIII
equation.

  Note that, even though
the perturbed weight  (\ref{weight-of-the-paper}) can be extended to a  $C^\infty(\mathbb{R})$ function when $t > 0$, it
has an essential singularity at the origin with respect to  complex variable $x$. In recent years, matrix models whose weight function has an essential singularity like \eqref{weight-of-the-paper} have appeared in several different areas of mathematics and physics; see, e.g.,   Berry and Shukla \cite{Berry:Shukla} in the study of statistics for zeros of the Riemann zeta function, Lukyanov \cite{Lukyanov} in a calculation of finite temperature expectation values in integrable quantum field theory, and \cite{Bro:Fra:Bee,Mez:Sim,Tex:Maj} in the study of the Wigner time delay in quantum transport. Wigner delay time stands for the average time that an electron spends when scattered by an open cavity and is of fundamental importance in the
theory of mesoscopic quantum dots. As suggested in \cite{Mez:Sim,Tex:Maj}, the distribution of
the Wigner delay time is far from being understood and several interesting questions remain
open.

In the Laguerre ensemble, the Wigner time delay is given by the sum of random variables $\tau_j$ such that
$1/\tau_j$ are distributed like the eigenvalues of matrices; see  \cite{Bro:Fra:Bee,  Tex:Maj}. The partition function, i.e., the quantity defined in \eqref{partition-function},   serves as the moment generating function of the probability density of the Wigner delay time \cite{Bri:Mez:Mo,Mez:Sim}.

Due to the influence of the essential singularity in the weight function, the asymptotic analysis of the above matrix models is very different from analytic cases as well as other singular cases, such as weights with jump discontinuities (see, e.g., \cite{ik,xz2011}), and weights with weak or algebraic  singularities (see, e.g. \cite{ik1, iko2009, xz2013b}). The first attempt to study asymptotics of matrix models with an essential singularity was done by Mezzadri and Mo \cite{Mez:Mo} and Brightmore  {\it{et al.}} \cite{Bri:Mez:Mo} when they are considering asymptotic properties of the partition function (the normalization constant $Z_n$ \eqref{partition-function}, in our notation) associated with the following weight
\begin{equation} \label{Mo-weight}
  w(x;z,s) = \exp\left( -\frac{z^2}{2x^2} + \frac{s}{x} - \frac{x^2}{2} \right), \quad z \in \mathbb{R} \setminus \{0\}, \ 0\leq s < \infty, \ x \in \mathbb{R}.
\end{equation}
Here $x=0$ is an essential singular point. As pointed out in \cite{Bri:Mez:Mo,Mez:Mo}, when $\alpha=\pm\frac{1}{2}$ and $s=0$, the system of polynomials orthogonal with respect to \eqref{weight-of-the-paper} and \eqref{Mo-weight} can be mapped to each other by a change of variables, however, the respective partition functions are still different. In \cite{Bri:Mez:Mo}, Brightmore  {\it{et al.}} showed that a phase transition emerges as the matrix size $n \to \infty$ and $s,z =O(1/\sqrt{n})$. They also obtained asymptotics of the partition function $Z_n$, which is characterized by a solution of a PIII equation. Although the PIII equation in \cite{Bri:Mez:Mo} is not the same as that in \eqref{v-eqn-introduction} below, similar phase transitions are observed; c.f. Theorems \ref{thm-limit-kernel}-\ref{thm-transition-to-Airy}. We think this may reflect  a new class of universality under the effect of the essential singularity. It is worth mentioning that, as in the current paper, the Deift-Zhou nonlinear steepest descent method is also used as one of the main tools in \cite{Bri:Mez:Mo}. The interested reader  may compare different model RH problems in Section \ref{sec:2} and \cite[Sec. 4.3]{Bri:Mez:Mo}, which are employed in the construction of local parametrices  near the essential singular points.

It is also interesting to study this problem from a polynomial point of view.
In fact, Chen and Its \cite{ci} use  (\ref{weight-of-the-paper}) as a  concrete and  important example of the Pollaczek-Szeg\"{o} type orthogonal polynomials, supported on infinite intervals. The Hankel determinant, which is the normalizing constant $Z_n$ in \eqref{partition-function}, plays a fundamental role in \cite{ci}, upon which the main results are derived and stated. A relation is also found between Hankel determinant and the Jimbo-Miwa-Ueno isomonodromy $\tau$-function.

For orthogonal polynomials with a certain weight $w(x)$ supported on $[-1, 1]$,  if the   Szeg\"{o} condition
\begin{equation*} \int^1_{-1} \frac {\ln w(x)}{\sqrt{1-x^2} }dx >-\infty \end{equation*} is fulfilled, then the weight is said to be of Szeg\"{o} class.
 The classical and the modified Jacobi weights belong to the    Szeg\"{o} class; cf. \cite{kmvv}. While the Pollaczek weight furnishes    a  well-studied  non-Szeg\"{o} class example, defined as
\begin{equation*} w(x; a, b)=\frac {e^{(2\theta-\pi)h(\theta)}}{\cosh [\pi h(\theta)]},~~\theta=\arccos x,~\mbox{for}~x\in (-1, 1),\end{equation*}
 where $h(\theta)=\frac {a\cos\theta +b} {2\sin\theta} =\frac {ax+b}{2\sqrt{1-x^2}}$, and $a$, $b$ are real constants such that $|b|<a$.  Asymptotics for the corresponding orthogonal polynomials can be found in    \cite[pp. 296-312]{s}; see also    \cite{zz2008}, where a Riemann-Hilbert analysis has been carried out.

It is readily seen  that   the  Pollaczek weight  just violates the Szeg\"{o} condition since
$$ w(x; a, b)\sim 2e^{a+b} e^{-\frac {C_+}{\sqrt{1-x}}}~~\mbox{as}~x\rightarrow 1^-,~~\mbox{and}~~w(x; a, b)\sim 2e^{a-b} e^{-\frac {C_-}{\sqrt{1+x}}}~~\mbox{as}~x\rightarrow -1^+,$$  where $C_\pm =\pi(a\pm b)/\sqrt 2>0$. In general, when the weight  behaves like $\exp\left \{-\frac C {(1-x)^\alpha}\right \}$   at an endpoint, say, $x=1$,
 with  constants   $\alpha \geq 1/2$ and  $C>0$, then the weight is of   non-Szeg\"{o} class.  Comparing \eqref{weight-of-the-paper}, we see that the weight $w(x;t)$ is of  non-Szeg\"{o} type at the origin.

The non-Szeg\"{o} class weights   demonstrate a singular behavior,  as compared with the classical polynomials; see Szeg\"{o} \cite[pp. 296-312]{s}.
 For example, such a singular behavior is shown in the extreme zeros.
  For the Pollaczek polynomials, the gap between the largest zero and the endpoint $1$ is of the order of magnitude $O(1/n)$, much bigger that the distance $O\left ( n^{-4/3}\right )$ between consecutive extreme zeros.
  This fact  is noticeable as compared with  the Jacobi polynomials: in the Jacobi case, both quantities are of the same order of magnitude $O(1/n^2)$.
   Hence, distinguishing the soft edge with the hard edge is not necessary in the Jacobi case. While it is
not so in the Pollaczek case. Such a singular behavior  is closely connected to the determination of the equilibrium measure, and to the `soft edge' appearing in later sections; cf. $\alpha_n$ in \eqref{kernel-Psi-Airy-approx}.

 It is worth mentioning that certain Szeg\"{o} class weights may also show a singular behavior, very similar to   the Pollaczek case; see  \cite{zxz2011} for an asymptotic analysis in such cases,  and see \cite{xzz2011} for an application in random matrix theory.

Now we can see from a polynomial point of view that for  $t>0$ fixed, the weight $w(x; t)$ in  (\ref{weight-of-the-paper}) is  of non-Szeg\"{o} class, and   the asymptotic behavior of the polynomials at the edge $x=0$ is expected to  be  described in terms of the Airy function, as  in   the  Pollaczek case; cf. \cite{zz2008}. While in the limiting case $t=0$, the weight in  (\ref{weight-of-the-paper}) reduces to the classical Laguerre weight, a typical  Szeg\"{o} class case, and the asymptotics  are described in terms of the Bessel functions; see \cite{kmvv, kv}. The really interesting piece here, might  be the Bessel to Airy transition, as the parameter $t$ shifts  from $t=0$ to a fixed positive number.   To achieve such a transition,  it is desirable to carry out a   large-$n$ asymptotic analysis for the orthogonal polynomials as $t=t_n\rightarrow 0$, or, eventually, uniformly for $t\in (0, d]$, with $d$ fixed and positive.
Note that, the Bessel to Airy transition in our paper is totally different from that in Claeys and Kuijlaars \cite{ck2008}, where the   PII functions appear.

In the same paper  \cite{ci}, along with an  investigation of the  Hankel determinants  and several relevant  statistic quantities,  an observation is   made, that for fixed degree $n$,  the corresponding polynomials are related to the  PIII equation   in a straightforward manner. Indeed,   Chen and Its \cite{ci} apply  the Riemann-Hilbert formulation of the orthogonal polynomials and the theory of Jimbo-Miwa,
to represent   the  polynomials orthogonal with respect to
  weight \eqref{weight-of-the-paper} via the Jimbo-Miwa Lax pair for the  PIII equation,  with parameters depending on the polynomial degree $n$.
Yet it is not easy to extract asymptotic approximations from such a PIII representation.
It is  noted in \cite{ci} that the asymptotics   of the polynomials for large  degree $n$ will provide valuable insight into the asymptotics of the Painlev\'{e} transcendent related.
However, to the best of our knowledge, there is no results on the asymptotics of the polynomials so far.

There are other examples where Painlev\'{e} equations  PI-PVI are involved. For instance, the three-term recurrence coefficients are shown related to PV in \cite{cd} for a Pollaczek-Jacobi type weight, and to
PIV in \cite{fvaz} for the semi-classic Laguerre weight.
Also, in  recent papers \cite{claeys-its-krasovsky2011, deift-its-krasovsky2011, krasovsky2011},   transition type uniform asymptotics of Hankel determinants have been considered.

The main objective   of the present  paper is to  obtain the limit behavior of the kernel $K_n(x,y)$  at the edge of the   spectrum for $t\in (0, d]$. Much attention will be paid to the transition of the edge behavior between  the Bessel   kernel and the Airy kernel, as   the parameter $t=t_n$ varies between  $t=0$  and  a fixed $d>0$. To achieve our goal, we  derive  the large degree  asymptotic behavior of the  orthogonal polynomials with respect to the weight (\ref{weight-of-the-paper}), uniformly for  $t=t_n\in (0, d]$, where $d$ is a positive constant. We use the Deift-Zhou nonlinear steepest descent method (also termed the Riemann-Hilbert approach) to   serve the purpose.

It is also of  interest to further consider, in a separate  paper,  several  asymptotic quantities such as  the Hankel determinant, the three-term recurrence  coefficients, the leading coefficients and extreme zeros of the corresponding orthogonal polynomials, in these quantities  the switch-on and switch-off  of a singular behavior   might  be observed, as the parameter $t$ varies in $(0,d]$.


\subsection{A third-order nonlinear differential equation and its reduction to PIII}

To state our results, we need to introduce a scalar function  $r(s)$ for $s\in [0, +\infty)$.   This function solves the following third-order nonlinear differential equation
\begin{equation} \label{r-eqn-introduction}
    2s^2 r' r''' - s^2 {r''}^2 + 2s r' r'' - 4  s {r'}^3 + \left(2r +2l- \frac{1}{4} \right) {r'}^2 + 1=0.
\end{equation}
The above equation is integrable. Its Lax pair is given as follows.
\begin{pro}\label{prop-r-equation}
    The equation \eqref{r-eqn-introduction} is the  compatibility condition $\Psi_{\zeta s} = \Psi_{s \zeta}$ of the following Lax pair
    \begin{eqnarray}
    \Psi_{\zeta}(\zeta, s) &=& \left(A_0(s) + \frac{A_1(s)}{\zeta} + \frac{A_2(s)}{\zeta^2} \right) \Psi(\zeta, s), \label{psi-lax-1} \\[.1cm]
    \Psi_{s}(\zeta, s) &=& \frac{B_1(s)}{\zeta} \Psi(\zeta, s), \label{psi-lax-2}
  \end{eqnarray}
  where
  \begin{equation} \label{psi-A's}
    A_0(s) =\left(
              \begin{array}{cc}
                0 & 0 \\
                \frac i 2 & 0 \\
              \end{array}
            \right)
       := \frac{i}{2} \sigma_-, \ A_1(s) = \begin{pmatrix}
    -\frac{1}{4} + \frac{1}{2}r(s) & -\frac{i}{2} \\
    -iq(s) & \frac{1}{4} - \frac{1}{2}r(s)
  \end{pmatrix}, \  A_2(s) = - s B_1(s).
  \end{equation}
  and
  \begin{equation} \label{psi-B}
    B_1(s) = \begin{pmatrix}
    q'(s) &-ir'(s) \\ it'(s) & -q'(s)
  \end{pmatrix}.
  \end{equation}
Here the functions $t(s)$ and $q(s)$ are determined in terms of $r(s)$ by
\begin{eqnarray}
 t'(s) &=& \frac{1 - q'(s)^2}{r'(s)}, \\
 q(s)  &=&   - s      r'(s) + \frac{1}{2} r(s)+\frac  1 2 r^2(s)+  l, \label{t-prime-q}
\end{eqnarray}
and $l$ is a constant.
\end{pro}


An observation shows that the   equation \eqref{r-eqn-introduction} can be reduced to a certain PIII equation. A proof will be given in Section \ref{sec:2.2}.

\begin{pro}\label{prop-PIII-reduction} By a change of unknown function
\begin{equation}\label{r-to-v}
  v(s)=sr'(s),
\end{equation}
the third-order equation \eqref{r-eqn-introduction} for $r(s)$ is reduced to a particular PIII equation for $v(s)$, namely,
  \begin{equation} \label{v-eqn-introduction}
    v''=\frac {{v'}^2}{v}-\frac {v'}{s}+\frac {v^2}{s^2} +\frac \alpha s-\frac 1 v;
  \end{equation}cf., e.g., \cite[(3.12)]{ci} for the PIII equation.
 \end{pro}

We note that the coefficients $l$ and $\alpha$, respectively in \eqref{r-eqn-introduction} and \eqref{v-eqn-introduction}, are determined by initial values of $r(s)$; see Section \ref{sec:2.2} below.
In the present paper, the constant $l$ in \eqref{r-eqn-introduction} and \eqref{t-prime-q} is equal to $0$.

We need a special solution of \eqref{r-eqn-introduction} which has no poles on $(0,+\infty)$.  Indeed, we have
\begin{pro} \label{prop-r-polefree}
 There exists a  solution   $r(s)$ of \eqref{r-eqn-introduction}, analytic for $s\in (0, +\infty)$, with the following boundary behaviors
 \begin{equation}
   r(0)=\frac 1 8\left (1-4\alpha^2\right ),~~\mbox{and}~~r(s)= \frac 32 s^{\frac 23}-\alpha s^{\frac 13}+O(1)~~\mbox{as}~~s\to +\infty.
 \end{equation}
\end{pro}

The existence of such an $r(s)$  follows directly from the vanishing lemma of the corresponding RH problem in Section \ref{sec:2.3} below. The initial value $r(0)$ is determined in \eqref{initial-condition}, and the asymptotic behavior at infinity is given in \eqref{r-asymptotics-at-plus-infty}.

\vskip .5cm

Given $r(s), q(s)$ and $t(s)$, the solutions of
\begin{equation} \label{psi-1-2-def}
  \frac{\partial}{\partial \zeta} \begin{pmatrix}
    \psi_1(\zeta, s) \\ \psi_2(\zeta, s)
  \end{pmatrix} = \left(A_0(s) + \frac{A_1(s)}{\zeta} + \frac{A_2(s)}{\zeta^2} \right) \begin{pmatrix}
    \psi_1(\zeta, s) \\ \psi_2(\zeta, s)
  \end{pmatrix}
\end{equation}
are analytic in the complex $\zeta$-plane with an essential singularity and a  possible branch point at $\zeta=0$. We define $\begin{pmatrix}
    \psi_1(\zeta, s) \\ \psi_2(\zeta, s)  \end{pmatrix}$ as the unique solution of the above equation with asymptotics
\begin{equation} \label{psi-1-2}
  \begin{pmatrix}
    \psi_1(\zeta, s) \\ \psi_2(\zeta, s)
  \end{pmatrix} = \biggl[ I + O(\zeta^{-1}) \biggr] \zeta^{-\frac{1}{4} \sigma_3} \frac{I + i \sigma_1}{\sqrt{2}} e^{\sqrt{\zeta} \sigma_3}  e^{\frac \pi 2 i(\alpha-1)  \sigma_3}
  \begin{pmatrix}
    1 \\ 1 \\
  \end{pmatrix}
    ~~\mbox{as}~\zeta \to \infty,
\end{equation}
and
\begin{equation} \label{psi-1-2-origin}
  \begin{pmatrix}
    \psi_1(\zeta, s) \\ \psi_2(\zeta, s)
  \end{pmatrix} =   e^{\frac \pi 2 i(\alpha-1) } \zeta^{\frac  \alpha 2} e^{\frac s \zeta} Q(s)  \left [ \begin{pmatrix}
    1 \\ 0
  \end{pmatrix}
   + O(\zeta)   \right ]    ~~\mbox{as}~\zeta \to 0.
\end{equation}
Both approximations are uniform in the sector $-\pi+\delta \leq \arg \zeta \leq -\frac{2\pi}{3} -\delta $ for small $\delta>0$.  In the above two formulas, the branches of powers of $\zeta$ are chosen such that
$\arg\zeta\in (-\pi, \pi)$,   $\sigma_1$ and $\sigma_3$ are the Pauli matrices; cf. \eqref{Pauli-matrix} below, and $Q(s)$ is independent of $\zeta$, such that $\det Q(s)=1$.  The uniqueness of the pair of functions is justified since they are recessive solutions at the origin in the above sector, so long as $s>0$; cf. \eqref{psi-1-2-origin}.     The functions $\psi_1(\zeta, s)$ and $\psi_2(\zeta, s)$ can be extended to  analytic functions in $\arg \zeta\in \mathbb{R}$, and will appear in our main results; see also  \eqref{psi1-psi2-def} and \eqref{Psi-jump}-\eqref{Psi-origin}  for an alternative definition. Note that they are well-defined for $s \in (0,+\infty)$ due to Proposition \ref{prop-r-polefree}.


\subsection{Main results}\label{sec:1.2}

Now we are ready to present out main results.

\subsection*{Limiting kernel at the hard edge}

The first main result is the $\Psi$-description of the limit of the re-scaled kernel $4n\, K_n(4nx, 4ny;t)$, where $K_n(x, y;t)$ is the polynomial kernel appeared in \eqref{kernel-CD-formula}, associated with the weight \eqref{weight-of-the-paper}. We focus on the large-$n$ behavior of the kernel near the  edge $x=0$:

{\thm{\label{thm-limit-kernel} Let $K_n(x,y;t)$ be the kernel given in \eqref{kernel-CD-formula}, then it has the   $\Psi$-kernel asymptotic approximation
\begin{equation}\label{kernel-approx-final-introduction}
\frac {1}{4n} K_n\left (\frac {u}{4n} ,\frac {v}{4n}; t \right )=K_\Psi(u,v, 2nt)+O\left (\frac 1 {n^2}\right)
 \end{equation}as $n\to\infty$,
uniformly for  $u$, $v$ in  compact subsets of  $(0,\infty)$ and  uniformly for  $t$ in $(0,d]$, where $d$ is a positive constant.
And the $\Psi$-kernel is given by
\begin{equation} \label{psi-kernel}
  K_\Psi(u,v,s)=\frac{  \psi_1(-v,s)\psi_2(-u,s)   -  \psi_1(-u,s)\psi_2(-v,s)}{2\pi i(u-v)}
\end{equation}
where the scalar function $\psi_k(\zeta, s)$, $k=1,2$,  are defined in \eqref{psi-1-2-def}-\eqref{psi-1-2-origin}}.
}

\vskip .5cm

Accordingly, the following result holds:

 {\cor{\label{thm-limit-kernel-double scailing limit} Let $K_n(x,y)$ be   the kernel given in \eqref{kernel-CD-formula}.
 If the parameter $t\to 0$ and $n\to \infty$ in the way such that
 $$\lim_{n\to\infty}2nt=\tau,\quad  \tau\in(0,\infty),$$
 we have the double scaling limit for  $K_n(x,y)$ given in terms of the $\Psi$-kernel defined in \eqref{psi-kernel}
 \begin{equation}
\lim_{n\to \infty} \frac {1}{4n} K_n\left (\frac {u}{4n} ,\frac {v}{4n}; t \right ) =K_\Psi(u,v,\tau)
 \end{equation}
uniformly  for $u$, $v$ and $\tau$  in  compact subsets of  $(0,\infty)$.
 }}

 {\rem{\label{scale-explain} The reader  may find it a little confusing to see the quantity $2nt$ on the right-hand-side of \eqref{kernel-approx-final-introduction}. The reason why we put \eqref{kernel-approx-final-introduction} in its current form is to describe the phase transition when the parameter $t$ varies in the interval $(0,d]$ (or, equivalently the parameter $s$ in the interval $(0,+\infty)$ if we let $s=2nt$). As one can see in the results here, when $t \sim \frac{1}{n}$ (i.e.  $s \sim 1$), we simply get the $\Psi$-kernel in \eqref{psi-kernel}; when $t = o(\frac{1}{n})$ (i.e. $s \to 0+$), the $\Psi$-kernel is reduced to the Bessel kernel in \eqref{bessel-kernel}; when $nt \to \infty$ (i.e. $s \to +\infty$) as $n \to \infty$, the $\Psi$-kernel is then reduced to the Airy kernel in \eqref{airy-kernel}.
}}

\medskip

 \subsection*{Transition to the Bessel kernel}

 The case dealt with in Theorem \ref{thm-limit-kernel} is for the parameter $s=2nt$ in compact subsets of $(0, \infty)$.
It is of interest to consider the   possible transition of the $\Psi$-kernel   in \eqref{kernel-approx-final-introduction} as $s\to 0^+$ and $s\to +\infty$. Indeed, by a nonlinear steepest descent analysis of the model RH problem for  small $s$, we have

 {\thm{\label{thm-transition-to-Bessel} We obtain the Bessel type limit for small parameter.
\begin{description}
\item (a)  The  $\Psi$-kernel is approximated by the Bessel kernel as $s\to 0^{+}$
   \begin{equation}\label{Psi-Bessel-approx-introduction}
 K_\Psi(u,v,s)= \mathbb{ J}_{\alpha}(u,v)+O(s),
\end{equation}
where   the error term is uniform for $u$ and $v$  in  compact subsets of  $(0,\infty)$.
The  Bessel kernel $\mathbb{J}_{\alpha}$ is defined in \eqref{bessel-kernel}.

 \item (b) If the parameter $t\to 0^{+}$ and $n\to \infty$ such that
 $$\lim_{n\to\infty}2nt=0,$$
 we have the Bessel kernel limit for $K_n$:
\begin{equation}\label{kernel-Bessel-approx-introduction}
 \lim_{n\to\infty} \frac {1}{4n} K_n\left (\frac {u}{4n} ,\frac {v}{4n}; t \right ) =\mathbb{ J}_{\alpha}(u,v),
  \end{equation}
uniformly  for $u$ and $v$  in  compact subsets of  $(0,\infty)$.
 \end{description}
 }   }

 \subsection*{Transition to the Airy kernel}

 It is even more interesting  to study  the asymptotic properties of the model RH problem for $\Psi(\zeta, s)$  when $s\to \infty$, and to describe the transition of the $\Psi$-kernel to the Airy kernel. Indeed, when the  $\Psi$ function is analyzed in the scale $\Psi(s^{2/3}\lambda, s)$ for large $s$, the behavior of it can be described at infinity in elementary functions, and at $\lambda=-1$ via the Airy function. From this fact we eventually obtain an Airy kernel limit as $s\to \infty$.  The results are summarized  in the following theorem.

  {\thm{\label{thm-transition-to-Airy} We obtain the Airy type limit for large parameter.
\begin{description}
\item (a)  The  $\Psi$-kernel is approximated by the Airy kernel as $s\to+\infty$
   \begin{equation}\label{Psi-Airy-approx-introduction}
\frac {s^{4/9}} c  K_\Psi\left (s^{2/3}\left (1-\frac {u}{cs^{2/9}}\right ),s^{2/3}\left (1-\frac {v}{cs^{2/9}}\right ),s\right )= \mathbb{ A}(u,v)+O\left (s^{- 2/9}\right ),
\end{equation}
where  the error term is uniform for $u$ and $v$  in  compact subsets of  $(-\infty,\infty)$, $c=(\frac 32)^{\frac 23}$ and the Airy kernel $\mathbb{A}$ is defined in \eqref{airy-kernel}.
 \item (b) If the parameter $t\in (0,d]$ and $n\to \infty$ such that
 $$\lim_{n\to\infty}2nt=\infty,$$
we have the Airy kernel limit for $K_n$:
\begin{equation}\label{kernel-Airy-approx-introduction}
\lim_{n \to \infty}\frac {\alpha_n}{cs^{2/9}} K_n\left (\alpha_n\left (1-\frac{u}{cs^{2/9}} \right ) ,\alpha_n\left ( 1-\frac{v}{cs^{2/9}}\right ); t \right )=\mathbb{ A}(u,v),
  \end{equation}
 where  $s=2nt$, $\alpha_n=s^{\frac 23}/(4n)=2^{-4/3}n^{-1/3}t^{ 2/3}$, $c=(\frac 32)^{\frac 23}$ and the formula  holds uniformly  for $u$ and $v$  in  compact subsets of  $(-\infty, \infty)$.
 \end{description}
 }   }

{\rem{\label{ MRS number} We note that the constant $\alpha_n$ appearing  in \eqref{kernel-Airy-approx-introduction} indicates the position of the   soft edge. Indeed, the equilibrium measure with potential  $\frac 1 n\left (x+\frac t{x}\right )$, $x\in(0,\infty)$ can be computed, and the support of the equilibrium measure turn out to be $(\alpha_n, 4n)$. From the perspective of Riemann-Hilbert approach to the universality of random matrices, the asymptotic behavior of the kernel $K_n$ at the soft edge is expected to be described in terms of the Airy kernel.
}}

\medskip

The rest of the paper is arranged as follows.
 In Section \ref{sec:2} we formulate  the model  RH problem for $\Psi(\zeta, s)$, prove its solvability for $x\in (0, \infty)$. We also derived a Lax pair
for $\Psi(\zeta,s)$, and show that the compatibility of the Lax pair leads to a third-order nonlinear
ordinary differential equation.
In Section \ref{sec:3}, we carry out, in full details,  the Riemann-Hilbert analysis  of the polynomials orthogonal with respect to  the weight functions    \eqref{weight-of-the-paper}.
    Section \ref{sec:4} will be devoted to the proof of   Theorem \ref{thm-limit-kernel}, based  on the asymptotic results  of the model problem $\Psi(\zeta, s)$  and of the RH problem associated with the weight \eqref{weight-of-the-paper}.
In   Section \ref{sec:5}, we investigate the $\Psi$-kernel to Bessel kernel transition and prove  Theorem \ref{thm-transition-to-Bessel}.
In the last section, Section \ref{sec:6}, we consider the $\Psi$-kernel to Airy kernel transition  and  prove Theorem \ref{thm-transition-to-Airy}.  Thus we complete the   Bessel to Airy transition
 as the parameter $t$ in \eqref{weight-of-the-paper} varies from left to right in a finite interval $(0, d]$.

\section{A model Riemann-Hilbert problem}
\setcounter{section} {2}
\setcounter{equation} {0} \label{sec:2}

\begin{figure}[h]
 \begin{center}
   \includegraphics[width=9cm]{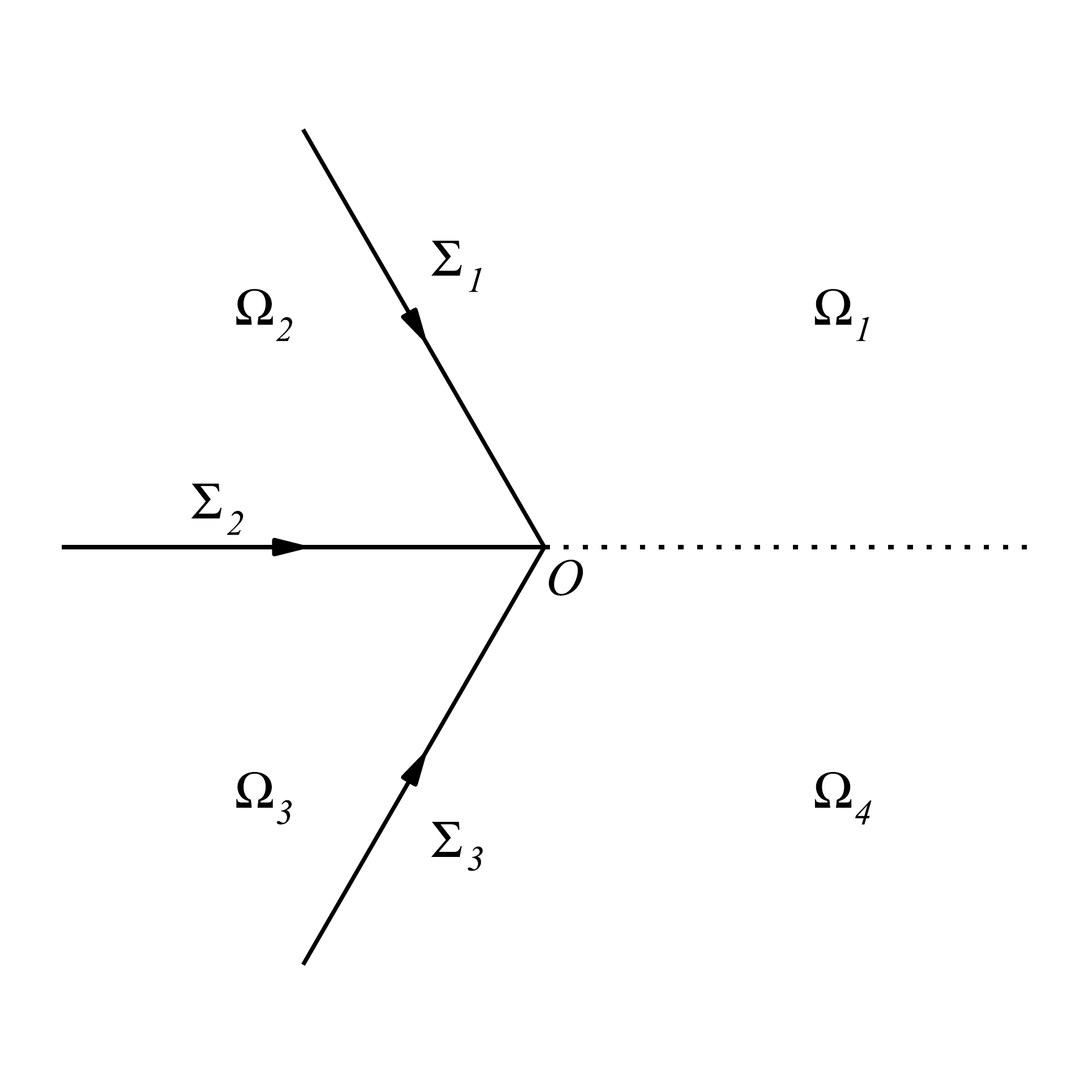} \end{center}
  \caption{\small{Contours and regions  for the model RH problem for $\Psi$  in the $\zeta$-plane, where both  sectors $\Omega_2$ and $\Omega_3$ have an opening angle  $\pi/3$.}}
 \label{contour-for-model}
    \end{figure}

We formulate a model RH problem, which will play a crucial role later in the steepest descent analysis. The model problem for $\Psi(\zeta)=\Psi(\zeta, s)$ is the following:

\begin{description}
  \item(a)~~  $\Psi(\zeta)$ is analytic in
  $\mathbb{C}\backslash\cup^3_{j=1}\Sigma_j$, where $\Sigma_j$   are illustrated in Figure \ref{contour-for-model};

  \item(b)~~  $\Psi(\zeta)$  satisfies the jump condition
 \begin{equation}\label{Psi-jump}
 \Psi_+(\zeta)=\Psi_-(\zeta)
 \left\{
 \begin{array}{ll}
    \left(
                               \begin{array}{cc}
                                 1 & 0 \\
                               e^{\pi i\alpha}& 1 \\
                                 \end{array}
                             \right), &  \zeta \in \Sigma_1, \\[.4cm]
    \left(
                               \begin{array}{cc}
                                0 &1\\
                               -1&0 \\
                                 \end{array}
                             \right),  &  \zeta \in \Sigma_2, \\[.4cm]
    \left(
                               \begin{array}{cc}
                                 1 &0 \\
                                  e^{-\pi i\alpha} &1 \\
                                 \end{array}
                             \right), &   \zeta \in \Sigma_3;
 \end{array}  \right .  \end{equation}

\item(c)~~  The asymptotic behavior of $\Psi(\zeta)$   at infinity
  is
  \begin{equation}\label{Psi-infty}
 \Psi(\zeta, s)= \left[ I + \frac{C_{1}(s)}{\zeta} +
O\left(\frac{1}{\zeta^2}\right) \right] \; \zeta^{-\frac{1}{4} \sigma_3} \frac{I + i \sigma_1}{\sqrt{2}} e^{\sqrt{\zeta} \sigma_3},~~\arg \zeta\in (-\pi, \pi),~~\zeta\rightarrow \infty,
   \end{equation}where $C_1(s)$ is a matrix independent of $\zeta$;

 \item(d)~~ The asymptotic behavior of $\Psi(\zeta)$  at $\zeta=0$ is
  \begin{equation}\label{Psi-origin}\Psi(\zeta,s)=Q(s)\left\{I+O(\zeta)\right\} e^{\frac s \zeta\sigma_3}\zeta^{\frac \alpha2\sigma_3}\left\{
  \begin{array}{ll}
    I, &  \zeta \in \Omega_1\cup\Omega_4, \\[.4cm]
   \left(
                               \begin{array}{cc}
                                 1 & 0 \\
                               -e^{\pi i\alpha}& 1 \\
                                 \end{array}
                             \right),  &  \zeta \in \Omega_2, \\[.4cm]
    \left(
                               \begin{array}{cc}
                                 1 &0 \\
                                  e^{-\pi i\alpha} &1 \\
                                 \end{array}
                             \right), &   \zeta \in \Omega_3
 \end{array}  \right .
  \end{equation} for $\arg \zeta\in (-\pi, \pi)$, as $\zeta\rightarrow 0$,
  where $\Omega_1-\Omega_4$ are depicted in Figure \ref{contour-for-model},  $Q(s)$ is a matrix independent of $\zeta$, such that $\det Q(s)=1$, and $\sigma_j$ are the Pauli matrices, namely,
  \begin{equation}\label{Pauli-matrix}
\sigma_1=\left(
                   \begin{array}{cc}
                     0 &1 \\
                    1 & 0 \\
                   \end{array}
                 \right),   ~~\sigma_2=\left(
                   \begin{array}{cc}
                     0 & -i \\
                    i & 0 \\
                   \end{array}
                 \right),~~\mbox{and}~\sigma_3=\left(
                   \begin{array}{cc}
                     1 & 0 \\
                    0 & -1 \\
                   \end{array}
                 \right).
 \end{equation}
 \end{description}

\subsection{Proof of Proposition \ref{prop-r-equation} and the Lax pair for $\Psi$}\label{sec:2.1}

In this subsection, we will prove Proposition \ref{prop-r-equation} and show that the above RH problem gives us the Lax pair given in \eqref{psi-lax-1} and \eqref{psi-lax-2}. The idea of the proof can be found in Fokas {\it{et al.}} \cite[Chap. 5]{fikn}.
\medskip

\noindent {\sc{Proof of Proposition \ref{prop-r-equation}}}.
Since  $\det \Psi(\zeta, s) \equiv 1$ by Liouville's theorem, we have $\textrm{tr} \, C_1(s) = 0$ in \eqref{Psi-infty}. Therefore,  we may denote
\begin{equation} \label{psi-c1-def}
  C_1(s) = \begin{pmatrix}
    q(s) & -ir(s) \\ it(s) & -q(s)
  \end{pmatrix}.
\end{equation}

All jumps in \eqref{Psi-jump}  are independent of $\zeta$ and $s$. Hence  we see that both $\Psi_\zeta \Psi^{-1}$ and  $\Psi_s \Psi^{-1}$ are  analytic in the whole complex $\zeta$-plane, except two possible isolated singularities at $\zeta=0$ and $\zeta=\infty$.

  From \eqref{Psi-infty}   and \eqref{Psi-origin}, it is easily verified that $\Psi_s \Psi^{-1}$ has a removable singularity  at $\infty$ and at most a  simple pole  at the origin. This gives us $\Psi_{s}= \left (B_0(s) + \frac{B_1(s)}{\zeta}\right )\Psi$. Moreover, a careful calculation from \eqref{Psi-infty}  yields
  \begin{eqnarray*}
    \Psi_s (\zeta) \Psi^{-1}(\zeta) \sim \frac{C_1'(s)}{\zeta}  \qquad \textrm{as } \zeta \to \infty,
  \end{eqnarray*}
  which implies that  $B_0(s)= 0$ and $B_1(s) = C_1'(s)$. Therefore, \eqref{psi-lax-2} follows. In addition, from \eqref{Psi-origin}  we have
  \begin{equation}
    \Psi_s (\zeta) \Psi^{-1}(\zeta) \sim \frac{1}{\zeta}Q(s) \sigma_3 Q(s)^{-1}   \qquad \textrm{as } \zeta \to 0.
  \end{equation}
  Hence we  obtain another representation for $B_1$, namely,  $B_1(s) = Q(s) \sigma_3 Q(s)^{-1}$, so that $\det C_1'(s)=\det B_1(s) = -1$, which in turn  gives
  \begin{equation} \label{B1-det}
    q'(s)^2 + r'(s)t'(s) = 1.
  \end{equation}

  Similarly, from \eqref{Psi-infty}   and \eqref{Psi-origin} we see that $\Psi_\zeta \Psi^{-1}$  has a removable singularity at infinity and a possible double pole at  $\zeta=0$. Writing
  $\Psi_{\zeta}(\zeta, s) = \left(A_0(s) + \frac{A_1(s)}{\zeta} + \frac{A_2(s)}{\zeta^2} \right) \Psi(\zeta, s)$,
  from \eqref{Psi-infty}
  we have
  \begin{eqnarray*}
    \Psi_\zeta (\zeta) \Psi^{-1}(\zeta) \sim \frac{i}{2} \sigma_- + \frac{1}{\zeta} \biggl(-\frac{1}{4} \sigma_3 + \frac{i}{2}[C_1, \sigma_-] - \frac{i}{2} \sigma_+ \biggr) \qquad \textrm{as } \zeta \to \infty,
  \end{eqnarray*}
  with $[A,B]=AB-BA$, and,  from \eqref{Psi-origin}
  \begin{eqnarray*}
    \Psi_\zeta (\zeta) \Psi^{-1}(\zeta) \sim - \frac{s}{\zeta^2} Q(s) \sigma_3 Q(s)^{-1} \qquad \textrm{as } \zeta \to 0.
  \end{eqnarray*}
  The  above two formulas give us \eqref{psi-lax-1}, with coefficient matrices given in \eqref{psi-A's}. Here use has been made of the  relation $A_2(s) = -s Q(s) \sigma_3 Q(s)^{-1}$.

  Next, we will derive the differential equation for $r(s)$ in \eqref{r-eqn-introduction} from the compatibility condition  of \eqref{psi-lax-1} and \eqref{psi-lax-2}. The compatibility condition, namely $\Psi_{\zeta s} =   \Psi_{s \zeta}$,  is now equivalent to
  \begin{equation*}
    \frac{\partial }{\partial s}\left(A_0(s) + \frac{A_1(s)}{\zeta} + \frac{A_2(s)}{\zeta^2} \right) - \frac{\partial }{\partial \zeta} \left(\frac{B_1(s)}{\zeta}\right) + \left[A_0(s) + \frac{A_1(s)}{\zeta} + \frac{A_2(s)}{\zeta^2}, \frac{B_1(s)}{\zeta} \right] = 0,
  \end{equation*}
  that is,
  \begin{equation*}
    A_0'(s) + \frac{A_1'(s) + [A_0(s), B_1(s)]}{\zeta} + \frac{A_2'(s) + B_1(s) + [A_1(s), B_1(s)]}{\zeta^2} + \frac{[A_2(s), B_1(s)]}{\zeta^3}=0.
  \end{equation*}
  Using \eqref{psi-B} and \eqref{psi-A's}, we see that only the $\zeta^{-2}$ term remains, and the last equation is thus reduced to
  \begin{equation*}
    -s C_1''(s) + [A_1(s), B_1(s)]  =0.
  \end{equation*}
  The above equation, together with \eqref{B1-det}, gives us the following equations for unknown scalar functions $q(s), r(s)$ and $t(s)$
  \begin{equation}\label{equation-sets}
    \begin{cases}
      s q''(s) =  q(s) r'(s) + \frac{1}{2}t'(s) \\[.1cm]
      s r''(s) = -q'(s) - \frac{1}{2}r'(s) +  r(s) r'(s) \\[.1cm]
      s t''(s) = -2  q(s) q'(s) + \frac{1}{2}t'(s) -   r(s)t'(s) \\[.1cm]
      q'(s)^2 + r'(s)t'(s) = 1.
    \end{cases}
  \end{equation}

  Note that, although there are 4 equations for 3 unknown functions, one of the equations is redundant. For example, the third equation in
  \eqref{equation-sets} can be deduced  from a combination of the first two equations, with the equation obtained by taking derivative  of the fourth equation, and hence the third equation may be removed.
  Now integrating the second equation in  \eqref{equation-sets} yields
 \begin{equation*}
     q(s)  =    - s      r'(s) + \frac{1}{2} r(s)+\frac  1 2 r^2(s)+  l,
  \end{equation*}where $l$ is a constant. In the present case, $l=0$, as can be seen from the initial conditions in \eqref{initial-condition}.
   Eliminating   $t(s)$  from the first and the last equation in  \eqref{equation-sets} gives
   \begin{equation*}
   -  sq''(s) r'(s) +\frac 1 2= -q(s) r'^2(s) +\frac 1 2 q'^2(s).
  \end{equation*}
   Further eliminating $q(s)$ from the last two equations yields   the following third-order nonlinear differential equation for $r(s)$
  \begin{equation} \label{r-eqn}
    2s^2 r' r''' - s^2 {r''}^2 + 2s r' r'' - 4  s {r'}^3 + \left(2 r +2l- \frac{1}{4} \right) {r'}^2 + 1=0,
  \end{equation}where the  constant  $l=0$.

\subsection{Proof of Proposition \ref{prop-PIII-reduction}: Reduction to PIII}\label{sec:2.2}
As is shown  in  Chen and Its \cite{ci},   various quantities, such as  the three-term recurrence coefficients of the associated orthogonal polynomials, are expressed in terms of a specific solution to a  PIII equation. On the other hand, in the present paper, the previous derivation   indicates that the third-order equation would play the same role. Hence, a reduction of \eqref{r-eqn-introduction} to  \eqref{v-eqn-introduction}   is  expected,  as established in Proposition \ref{prop-PIII-reduction}.

 Now  we  prove Proposition \ref{prop-PIII-reduction} by showing \eqref{r-eqn-introduction} and \eqref{v-eqn-introduction} are equivalent.

\medskip

\subsubsection*{Reduction \eqref{v-eqn-introduction}$\ \Longrightarrow\ $\eqref{r-eqn-introduction} }

\medskip

Substituting  $v=sr'$ into \eqref{v-eqn-introduction}, we obtain an alternative third-order equation
\begin{equation}\label{formal asy}
s^2r'r'''-s^2r''^2+sr'r''-s r'^3-\alpha r'+1=0.\end{equation}
Multiplying $2r''/r'^3$, and grouping the terms, we can put \eqref{formal asy} into the   form  of a differential
\begin{equation} \label{r-differential}
 \frac{d}{ds}\left\{\frac{s^2 {r''}^2}{{r'}^2}- 2  s r'+\frac {2\alpha}{r'}-\frac 1{{r'}^2}+2r\right \}=0.
  \end{equation}
Integrating \eqref{r-differential} yields
\begin{equation} \label{r-order 2-1}
 {s^2 {r''}^2}- 2 s r'^3+  {2\alpha}{r'}-  1 +2r r'^2=\left ( \frac{1}{4}-2l\right ){{r'}^2},
  \end{equation}
with an integral constant
\begin{equation} \label{constant l}
l=\frac 18-r(0)-\frac {\alpha} {r'(0)}+\frac {1}{2r'(0)^2}.
  \end{equation}
Multiplying \eqref{formal asy} by 2 and adding it to \eqref{r-order 2-1} give us \eqref{r-eqn-introduction}.  Here, we note that in the present case,
  with the special initial values $r(0)=\frac 1 8\left (1-4\alpha^2\right )$ and $r'(0)= \frac 1 \alpha$, the constant $l$ in \eqref{constant l} vanishes.

\medskip

\subsubsection*{Reduction \eqref{r-eqn-introduction}$\ \Longrightarrow\ $\eqref{v-eqn-introduction}}

\medskip

Reversely, we proceed to show that  \eqref{r-eqn-introduction} implies \eqref{formal asy}. To this aim, we denote by $\Lambda$  the left hand side of \eqref{formal asy}.
In view of the linear dependence of \eqref{r-eqn}, \eqref{formal asy} and \eqref{r-order 2-1} mentioned above, paying attention to the equivalence of \eqref{formal asy} and \eqref{r-differential},  we see that
\begin{equation} \label{equation for Lambda}
\left(\frac {2 }{r'^2}\Lambda\right)'=-\frac {2r''}{r'^3}\Lambda.
  \end{equation}
Solving the equation, we have
 \begin{equation} \label{equation for Lambda-solution}
\Lambda=\alpha_1 r', \quad \alpha_1=\alpha-\frac 1{r'(0)},
  \end{equation}
 where   the constant $\alpha_1$ is determined by comparing  both sides at $s=0$. For the chosen initial value $r'(0)=\frac 1 \alpha$, the constant  vanishes,
thus  we get  $\Lambda=0$, which is the third-order equation \eqref{formal asy}.
Finally, substituting   $v=s r'$, or, equivalently, $r'=v/s$, into \eqref{formal asy},  we obtain the  PIII equation \eqref{v-eqn-introduction}.

\subsection{Solvability of the model Riemann-Hilbert problem}\label{sec:2.3}

 We proceed to justify  the solvability of the RH problem   for $\Psi(\zeta, s)$,  by proving  a vanishing lemma.
{\lem{\label{solvability}Assume that the homogeneous RH problem for
$\Psi^{(1)}(\zeta, s)$ adapting   the same jump conditions
(\ref{Psi-jump})  and the same boundary condition
(\ref{Psi-origin}) as $\Psi(\zeta, s)$,
 with the behavior (\ref{Psi-infty})  at infinity
   being altered  to
 \begin{equation}\label{Psi-infty-homogeneous}\Psi^{(1)}(\zeta, s)=O\left (\frac 1 \zeta\right )
  \zeta^{-\frac{1}{4}\sigma_3}\frac{I+i\sigma_1}{\sqrt{2}}
  e^{\sqrt{\zeta}\sigma_3},~~\arg \zeta\in (-\pi, \pi),~~\zeta\rightarrow \infty.\end{equation}
  If the parameter $s\in (0, +\infty)$,
   then   $\Psi(\zeta, s)$ is trivial, that is, $\Psi\equiv 0$.
}}\vskip .2cm

\noindent {\sc{Proof}}. First, we remove the exponential factor  at
infinity and eliminate the jumps on $\Sigma_1$ and
$\Sigma_3$ by defining
\begin{equation}\label{Psi2-def}
\Psi^{(2)}(\zeta)=\left \{
\begin{array}{ll}
  \Psi^{(1)}(\zeta)e^{-\sqrt{\zeta}\sigma_3}, &  \zeta\in\Omega_1\cup\Omega_4,
  \\[.3cm]
  \Psi^{(1)}(\zeta)e^{-\sqrt{\zeta}\sigma_3} \left( \begin{array}{cc}
                                 1 & 0 \\
                                  e^{\pi i\alpha}e^{-2\sqrt{\zeta}} & 1 \\
                               \end{array}
                             \right), &  \zeta\in\Omega_2,
                             \\[.3cm]
\Psi^{(1)}(\zeta)e^{-\sqrt{\zeta}\sigma_3} \left( \begin{array}{cc}
                                 1 & 0 \\
                                  -e^{-\pi i\alpha}e^{-2\sqrt{\zeta}} & 1 \\
                               \end{array}
                             \right), &  \zeta\in\Omega_3;
\end{array}\right.\end{equation}
cf.  Figure \ref{contour-for-model} for the  regions $\Omega_1-\Omega_4$, where $\arg\zeta\in (-\pi, \pi)$.

It is easily verified  that $\Psi^{(2)}(\zeta)$ solves the
following RH problem:
\begin{description}
  \item(a)~~  $\Psi^{(2)}(\zeta)$  is analytic in
  $\zeta\in \mathbb{C}\backslash{\Sigma}_2$ (see Figure \ref{contour-for-model});

  \item(b)~~  $\Psi^{(2)}(\zeta)$  satisfies the jump condition
   \begin{equation}\label{Psi2-jump}
  \left(\Psi^{(2)}\right )_+(\zeta)=\left(\Psi^{(2)}\right )_-(\zeta)\left(
                              \begin{array}{cc}
                                 e^{-(2\sqrt{\zeta}_{+}-\pi i \alpha)} & 1 \\
                                  0 &  e^{(2\sqrt{\zeta}_+-\pi i\alpha)} \\
                               \end{array}\right), ~~\zeta \in \Sigma_2, \end{equation}
where $\arg \zeta\in(-\pi,\pi)$, and $\sqrt{\zeta}_{+}=i\sqrt{|\zeta|}$ for $\zeta\in  \Sigma_2$;
\item(c)~~  The asymptotic behavior of $\Psi^{(2)}(\zeta)$  at infinity
  is
  \begin{equation}\label{Psi2-infty}\Psi^{(2)}(\zeta)=O\left(\zeta^{-\frac 34}\right),~~\arg \zeta\in (-\pi, \pi),~~\zeta\rightarrow\infty;\end{equation}
\item(d)~~The behavior of $\Psi^{(2)}(\zeta)$  at the origin is
\begin{equation}\label{Psi2-origin}
\Psi^{(2)}(\zeta)= O(1) e^{\frac s \zeta  \sigma_3}\zeta^{\frac \alpha2\sigma_3},~~\arg \zeta\in (-\pi, \pi),~~\zeta\rightarrow 0.
\end{equation}
\end{description}

{\rem{\label{rem-x-great-than-0}
It is worth  noting that a consistency check of (\ref{Psi2-jump}) and (\ref{Psi2-origin})  yields  $s>0$. Indeed,   it is seen that the $O(1)$ factor in (\ref{Psi2-origin}) stands for an invertible matrix, bounded and with determinant $1$, and thus having an $O(1)$ inverse. Substituting
(\ref{Psi2-origin}) into (\ref{Psi2-jump}) leads eventually to $\left(
                                                                  \begin{array}{cc}
                                                                    e^{-2\sqrt{\zeta_+}} & |\zeta|^\alpha e^{-2s/|\zeta|}  \\
                                                                    0 & e^{2\sqrt{\zeta_+}} \\
                                                                  \end{array}
                                                                \right)=O(1)$, and $s\in (-\infty, 0)$ is thus  declined. Instead, we consider the solvability of $\Psi(\zeta, s)$ and the analyticity in $s$ only for $s>0$.
                                                                           }}\vskip .5cm

We carry out yet another transformation to move the oscillating entries  in the jump matrices to off-diagonal, as follows:
\begin{equation}\label{Psi3-def}
\Psi^{(3)}(\zeta)=\left \{
\begin{array}{ll}
  \Psi^{(2)}(\zeta) \left( \begin{array}{cc}
                                 0 & -1 \\
                                  1 & 0 \\
                               \end{array}
                             \right) , & \mbox{for}~ \Im\zeta>0,
                              \\[.2cm]
\Psi^{(2)}(\zeta), & \mbox{for}~ \Im\zeta<0.
\end{array}\right .\end{equation}
Then $\Psi^{(3)}(\zeta)$ solves  a RH problem  with   jumps
\begin{equation}\label{Psi3-jump-formula}
\left (\Psi^{(3)}\right )_+(\zeta)=\left (\Psi^{(3)}\right
)_-(\zeta)J^{(3)}(\zeta),~~\zeta\in \mathbb{R},\end{equation}
where
\begin{equation}\label{Psi3-jump}J^{(3)}(\zeta)=
\left\{\begin{array}{ll}
        \left(
                               \begin{array}{cc}
                                  1& - e^{-(2\sqrt{\zeta}_{+}-\pi i \alpha)}\\
                                  e^{(2\sqrt{\zeta}_{+}-\pi i \alpha)} & 0 \\
                                 \end{array}
                             \right),  &  \zeta \in(-\infty,0), \\[.5cm]
          \left( \begin{array}{cc}
                                 0 & -1 \\
                                  1 & 0 \\
                               \end{array}
                             \right),& \zeta \in(0, \infty);
       \end{array}\right . \end{equation}
Furthermore, the behavior of $\Psi^{(3)}$  at infinity is still of the form  (\ref{Psi2-infty}),      while  the   condition at $\zeta= 0$ now takes
\begin{equation}\label{Psi3-origin}
\Psi^{(3)}(\zeta)\sigma_2 =O(1)e^{\frac s \zeta  \sigma_3}\zeta^{\frac \alpha2\sigma_3},~\Im \zeta>0,~~\mbox{and} ~~ \Psi^{(3)}(\zeta)=O(1)e^{\frac s \zeta \sigma_3}\zeta^{\frac \alpha2\sigma_3},~\Im \zeta<0,
\end{equation}
recalling that $-i\sigma_2=\left(
                    \begin{array}{cc}
                      0 & -1 \\
                      1& 0 \\
                    \end{array}
                  \right)
$.

 It is readily seen that
\begin{equation}\label{Psi3-conjugate-sum}
(J^{(3)}(\zeta))^{*}+J^{(3)}(\zeta)=2\left(
                                            \begin{array}{cc}
                                              1 & 0 \\
                                             0 & 0 \\
                                            \end{array}
                                          \right),~~\zeta \in (-\infty, 0),
\end{equation}
where $X^*$ denotes the Hermitian conjugate of the matrix $X$.

 Next, we define an auxiliary matrix function
 \begin{equation}\label{Psi3-H-def}
H(\zeta)=\Psi^{(3)}(\zeta) \left ( \Psi^{(3)}(\bar \zeta\
)\right )^* , \quad \zeta\not \in \mathbb{R}.
\end{equation}
Then $H(\zeta)$
 is analytic in $\mathbb{C}\backslash \mathbb{R}$.  Substituting (\ref{Psi2-infty}) and (\ref{Psi3-origin}) to (\ref{Psi3-H-def}) gives
\begin{equation*} H(\zeta)=O\left ( \zeta^{-\frac 3 2}\right )~~\mbox{as}~~\zeta\rightarrow\infty, \end{equation*}
and
\begin{equation*} H(\zeta)=O(1)~~\mbox{as}~~\zeta\rightarrow 0.
  \end{equation*}

Thus, by Cauchy's integral  formula, we have
\begin{equation}\label{Psi3-H-Cauchy}\int_{\mathbb{R}}H_+(\zeta)d\zeta=0.
  \end{equation}
Now in view of  (\ref{Psi3-conjugate-sum}), and adding  to (\ref{Psi3-H-Cauchy}) its
Hermitian conjugate, we have
\begin{equation*}
2\int_{-\infty}^0\left (\Psi^{(3)}\right )_-(\zeta) \left(
                               \begin{array}{cc}
                                1 & 0 \\
                                 0 & 0 \\
                                 \end{array}
                             \right)
\left (\Psi^{(3)}\right )_-^*(\zeta) d\zeta= 0.
\end{equation*}

A straightforward  consequence is that  the first column of $\left(\Psi^{(3)}\right)_-(\zeta)$ vanishes for $\zeta\in (-\infty,0)$.
 Furthermore, it follows from \eqref{Psi3-jump-formula}   that the second column of $\left(\Psi^{(3)}\right)_{+}(\zeta)$ vanishes, also for  $\zeta\in (-\infty,0)$.

The jump $J^{(3)}(\zeta)$ in (\ref{Psi3-jump}) admits an analytic continuation in a neighborhood of $(-\infty,0)$. Accordingly,
$$\tilde\Psi^{(3)}(\zeta):=\left\{\begin{array}{ll}
                            \Psi^{(3)}(\zeta), & \arg\zeta\in (0, \pi), \\ [.3cm]
                           \Psi^{(3)}(\zeta e^{-2\pi i})\left(
                               \begin{array}{cc}
                                  1& - e^{-(2\sqrt{\zeta} -\pi i \alpha)}\\
                                  e^{(2\sqrt{\zeta} -\pi i \alpha)} & 0 \\
                                 \end{array}
                             \right),
                             & \arg\zeta\in (\pi, 2\pi)
                          \end{array}\right .$$
defines an analytic function in the cut-plane $\arg\zeta\in (0, 2\pi)$, such that $\tilde\Psi^{(3)}(\zeta)=\Psi^{(3)}(\zeta)$ for $\Im\zeta>0$, and $\tilde\Psi^{(3)}(\zeta)=\left (\Psi^{(3)}\right )_+(\zeta)$
                           for $\zeta\in (-\infty, 0)$. Hence we have
\begin{equation}\label{Psi3-2nd-column-upper}       \left (\Psi^{(3)}\right )_{12}(\zeta) = \left (\Psi^{(3)}\right )_{22}(\zeta)  = 0,~~\Im\zeta >0.\end{equation}

Similarly, we can also obtain
\begin{equation}\label{Psi3-1st-column-lower}      \left (\Psi^{(3)}\right )_{11}(\zeta) = \left (\Psi^{(3)}\right )_{21}(\zeta)  = 0,~~\Im\zeta <0.\end{equation}
The reader is referred to \cite{xz2011} and \cite{xz2013b} for a similar argument.

Now we proceed to exam the other entries of
$\Psi^{(3)}(\zeta)$ by appealing to  Carlson's
theorem (cf. \cite[p.236]{rs}). To this aim, for $k=1,2$,  we define  scalar
functions
\begin{equation}\label{Psi3-gk-def}
g_k(\zeta)=\left \{
\begin{array}{ll}\left(
 \Psi^{(3)}(\zeta)\right)_{k1},
 ~\mbox{for}~0<\arg \zeta <\pi, \\[.2cm]
\left(
 \Psi^{(3)}(\zeta)\right)_{k2},
 ~\mbox{for}~-\pi<\arg\zeta<0.
\end{array}\right .\end{equation}
From    (\ref{Psi3-jump}) and \eqref{Psi3-2nd-column-upper}-(\ref{Psi3-gk-def}),  we see  that
 each  $g_k(\zeta)$ is analytic in
  $\mathbb{C}\backslash (-\infty, 0]$,  and satisfies
  the jump conditions
  \begin{equation}\label{Psi3-gk-jump}
  \left (g_k\right )_+(\zeta)=\left (g_k\right )_-(\zeta)e^{2\sqrt{\zeta}_{+}-\pi i\alpha } , \quad \zeta \in (-\infty,0).
  \end{equation}

The sector of
  analyticity of $g_k(\zeta)$ can be extended as  follows:
\begin{equation}\label{Psi3-gk-ext}
 \hat{g}_k(\zeta)=\left \{
\begin{array}{ll} g_k(e^{-2\pi i}\zeta ) e^{-\pi i\alpha}e^{2\sqrt{\zeta}},
\quad &
 \mbox{for}~  \pi \leq\arg \zeta <2\pi, \\
g_k(e^{2\pi i}\zeta ) e^{\pi i\alpha}e^{2\sqrt{\zeta}},\quad
&
 \mbox{for}~  -2\pi<\arg\zeta\leq-\pi.
\end{array}\right .
\end{equation}
Thus $\hat{g}_k(\zeta)$ is now analytic in a  sector $-2\pi<\arg \zeta<2\pi$.
It is worth noting that the function $\hat{g}_k(\zeta)$,  so defined, is actually   analytic in a larger sector $-3\pi<\arg\zeta <3\pi$, and  the exponential term $|e^{ \sqrt{\zeta}}|\leq1$ for $\pi \leq\arg \zeta \leq 2\pi$ and $-2\pi\leq \arg\zeta\leq-\pi$.

If we put
\begin{equation}\label{Psi3-hk-def}
 h_k(\zeta)=\hat{g}_k((\zeta+1)^4)  \quad \mbox{for} \ \arg\zeta\in [-\pi/2,
 \pi/2],
\end{equation}
then the above discussion implies that $h_k(\zeta)$ is analytic in
$\Re \zeta >0$, continuous and bounded in  $\Re\zeta\geq 0$, and
satisfies the  decay condition on the imaginary axis
\begin{equation}\label{Psi3-hk-decay}
 |h_k(\zeta)|=O\left ( e^{- |\zeta|^2}\right ), \quad \mbox{for} \ \Re \zeta=0~ \mbox{as}~ |\zeta|\rightarrow \infty.
\end{equation}  Hence   Carlson's
theorem applies, and we have $h_k(\zeta)\equiv 0$ for
$\Re \zeta>0$. Tracing back, we see that all entries  of $\Psi^{(3)}(\zeta)$
vanish for $\zeta\not\in \mathbb{R}$. Therefore,  $\Psi^{(3)}(\zeta)$ vanishes identically, which implies that  $\Psi^{(1)}(\zeta)$ vanishes identically. This completes the proof of the vanishing lemma.\hfill\qed\\

The solvability of  the RH problem for $\Psi_0$ follows from the   vanishing lemma. As briefly indicated in \cite[p.104]{fikn}, the RH problem is equivalent to a Cauchy-type
singular integral equations, the corresponding singular integral operator is a Fredholm operator of index zero. The vanishing
lemma states that the null space is trivial, which implies that the singular integral equation (and thus $\Psi_0$) is solvable as a result of the
Fredholm alternative theorem. More details can be found in \cite[Proposition 2.4]{ik1}; see also \cite{deift,dkmv1,fikn,fz} for  standard methods connecting RH problems with integral equations.

Now we have the following solvability result:

{\lem { For $s\in (0, \infty)$, there exists a unique solution to the  RH problem
(\ref{Psi-jump})-(\ref{Psi-origin}) for $\Psi(\zeta,s)$. }}

\section{ Nonlinear steepest descent analysis } \indent\setcounter{section} {3}
\setcounter{equation} {0} \label{sec:3}

\noindent
This whole section  will be devoted to the asymptotic analysis of the orthogonal polynomials with respect to the weight $w(x;t)$ given in (\ref{weight-of-the-paper}).
We begin with  a RH formulation $Y(z)$ of the orthogonal polynomials.
Such a  remarkable           connection between the orthogonal polynomials and
 RH problems is observed by    Fokas, Its and Kitaev \cite{fik}.
  Then, we apply the nonlinear
steepest descent analysis  developed by Deift and Zhou {\it{et al.}}
\cite{dkmv1,dkmv2} to the RH problem for $Y$; see also  Bleher and  Its \cite{bi}.
The idea is to obtain,
via a series of invertible transformations $Y
\rightarrow T \rightarrow S \rightarrow R$, eventually the  RH problem for $R$, with  jumps   close to the identity
matrix, where
\begin{itemize}
\item $Y\to T$ is to re-scale the variable, to  accomplish  a normalization of $Y(z)$ at infinity, and to remove the exponential factor $e^{-t/x}$  in  the weight function $w$. As a result, $T(z)$ solves an oscillatory RH problem, normalized at infinity.
  \item $T\to S$ is based on a factorization of the oscillatory jump, and a deformation of the contours. $S(z)$ solves a RH problem without oscillation, yet the contours are  self-intersected.
  \item $S\to R$, the final transformation, leads to a RH problem for $R(z)$ with all jumps close to $I$, and $R(z)$ can then be expanded on the whole complex plane into a Neumann series. We use only the leading term in the present paper, though. To apply the transformation, a parametrix at the outside region, and local  parametrices  at the origin and at the soft edge $z=1$ have to be constructed.
  \end {itemize}
 Tracing back,   the uniform asymptotics of the orthogonal
polynomials in the complex plane is obtained for large polynomial degree  $n$.
Technique difficulties lie in   the construction of the  local parametrix in a neighborhood of the
origin $z=0$. The parametrix   possesses   irregular singularity both at infinity and at the origin.

\subsection{  Riemann-Hilbert problem for orthogonal polynomials }
Initially,  the RH problem for orthogonal polynomials
is as follows (cf. \cite{fik}).
\begin{description}
  \item(Y1)~~  $Y(z)$ is analytic in
  $\mathbb{C}\backslash [0,\infty)$;

  \item(Y2)~~  $Y(z)$  satisfies the jump condition
  \begin{equation}\label{Y-jump}
  Y_+(x)=Y_-(x) \left(
                               \begin{array}{cc}
                                 1 & w(x) \\
                                 0 & 1 \\
                                 \end{array}
                             \right),
\qquad x\in (0,\infty),\end{equation} where $w(x)=w(x;t)=x^\alpha e^{-x-t/x}$ is the
weight function defined  in
 (\ref{weight-of-the-paper});
  \item(Y3)~~  The asymptotic behavior of $Y(z)$  at infinity is
  \begin{equation}\label{Y-infty}Y(z)=\left (I+O\left (  1 /z\right )\right )\left(
                               \begin{array}{cc}
                                 z^n & 0 \\
                                 0 & z^{-n} \\
                               \end{array}
                             \right),\quad \mbox{as}\quad z\rightarrow
                             \infty ;\end{equation}
\item(Y4)~~The asymptotic behavior of $Y(z)$   at the end points $z=0$ are
 \begin{equation}\label{Y-origin}Y(z)=\left(
                               \begin{array}{cc}
                                O( 1) &  O( 1) \\[0.2cm]
                                O( 1) & O( 1)
                                  \\
                               \end{array}
                             \right),\quad \mbox{as}\quad z\rightarrow
                             0 .\end{equation}

\end{description}
\vskip .5cm

 By virtue of the Plemelj formula and Liouville's theorem, it is known that the above RH problem for $Y$
 has a unique solution
\begin{equation}\label{Y-solution}
Y(z)= \left (\begin{array}{cc}
\pi_n(z)& \frac 1 {2\pi i}
\int_{0} ^{\infty}\frac {\pi_n(s) w(s) }{s-z} ds\\[0.2cm]
-2\pi i \gamma_{n-1}^2 \;\pi_{n-1}(z)& -   \gamma_{n-1}^2\;
\int_{0} ^{\infty}\frac {\pi_{n-1}(s) w(s) }{s-z} ds \end{array} \right ),
\end{equation}
where  $\pi_n(z)$ is the monic polynomial, and $p_n(z)=\gamma_{n}\pi_n(z)$
is the  orthonormal  polynomial with respect to the weight
$w(x)=w(x;t)$;   cf., e.g.,  \cite{deift} and \cite{fik}.\\
\vskip .3cm

\noindent

\subsection{ The first transformation  $Y\rightarrow T$ }\label{sec:3.2}
The first transformation is to normalize the above RH problem for
$Y$ at infinity. Beforehand, we write down the equilibrium measure with
the external field $V(x)=4x,x>0$, that is,
\begin{equation}\label{psi-e-measure }\psi(x)=\frac 2{\pi}\sqrt{\frac{1-x}{x}},\quad 0<x<1;\end{equation}  cf. \cite{vanlessen}, see also \cite{qw}.
For later use,  we defined several other auxiliary functions
\begin{equation}\label{g-function}g(z)=\int_0^1  \ln(z-x) \psi(x) dx, \end{equation}
where the branch is chosen such that $\arg(z-x)\in(-\pi,\pi)$, and
\begin{equation}\label{phi-function}\phi(z)=2 \int_0^z\sqrt{\frac{s-1}{s}}ds,~~z\in
\mathbb{C}\backslash[0,\infty),
\end{equation}  where $\arg z\in(0,2\pi)$, such that the Maclaurin expansion $\phi(z)=  4i\sqrt{z}\left\{ 1-\frac z 6+\cdots  \right \}$ holds for $|z|<1$.

The first transformation  $Y\rightarrow T$ is defined as
\begin{equation}\label{TrsnaformationY-T}T(z)= (4n)^{-(n+\frac \alpha 2) \sigma_3}e^{-n\frac{1}{2}l\sigma_3}Y(4nz) e^{-n(g(z)-\frac{1}{2}l)\sigma_3}e^{-\frac {t_n}{8nz}\sigma_3} (4n)^{\frac \alpha 2\sigma_3} \end{equation} for $z\in
\mathbb{C}\backslash[0,\infty)$, where $l=-2(1+\ln 4)$ is the Euler-Lagrange
constant. The purpose of the transformation is threefold: to re-scale the variable, to  accomplish  a normalization of $Y(z)$ at infinity, and to remove the exponential factor in  the weight function $w$. Here $t=t_n$ indicates the dependence of the    parameter $t$ on $n$, the polynomial degree.
Then $T$ solves the RH problem:
\begin{description}
\item(T1)~~ $T(z)$ is analytic in
$\mathbb{C}\backslash [0,\infty)$;
\item(T2)~~  The jump condition is
\begin{equation}\label{T-jump}T_+(x)=T_-(x)
\left(
                               \begin{array}{cc}
                                 e^{n(g_ -(x)-g_+(x))} & x^{\alpha}e^{n(-4x+g_+(x)+g_-(x)-l)} \\
                                 0 & e^{n(g_ +(x)-g_-(x))} \\
                                 \end{array}
                             \right),\quad x \in (0, \infty);
\end{equation}
\item(T3)~~   The asymptotic behavior of $T(z)$  at infinity is
\begin{equation}\label{T-infty}T(z)= I+O(1/z)\quad \mbox{as}\quad z\rightarrow \infty ;\end{equation}
\item(T4)~~  The asymptotic behavior of $T(z)$    at the end points  $z=0$ is
\begin{equation}\label{T-origin}T(z)=O(1)e^{-\frac {t_n}{8nz}\sigma_3}.\end{equation}
\end{description}

From (\ref{g-function}) and (\ref{phi-function}) it is readily seen that
$$g_+(x)-g_-(x)=2\pi i-2\phi_+(x),~~x\in (0, 1).$$
Also, one of the
phase conditions reads
$$  -4x+g_+(x)+g_-(x)-l=0,~~ \mbox{for}~x\in(0,1),
$$ with  the Euler-Lagrange constant involved.
Hence, the jumps in (\ref{T-jump}) can be represented in $\phi$, as follows:
\begin{equation}\label{T-jump-in-phi}T_+(x)=T_-(x)\left\{
\begin{array}{ll}
 \left(
                               \begin{array}{cc}
                                 1 & x^{\alpha}e^{-2n \phi(x)} \\
                                 0 & 1 \\
                               \end{array}
                             \right),&   x\in(1,\infty);\\[.4cm]
                                \left(
                               \begin{array}{cc}
                                e^{2n\phi_ +(z)}& x^{\alpha} \\
                                 0 &e^{2n\phi_ -(z)} \\
                                 \end{array}
                             \right), &  x\in(0,1).
\end{array}
\right .
\end{equation}

\subsection{ The second  transformation  $T\longrightarrow S$}
\begin{figure}[t]
 \begin{center}
\includegraphics[width=9cm]{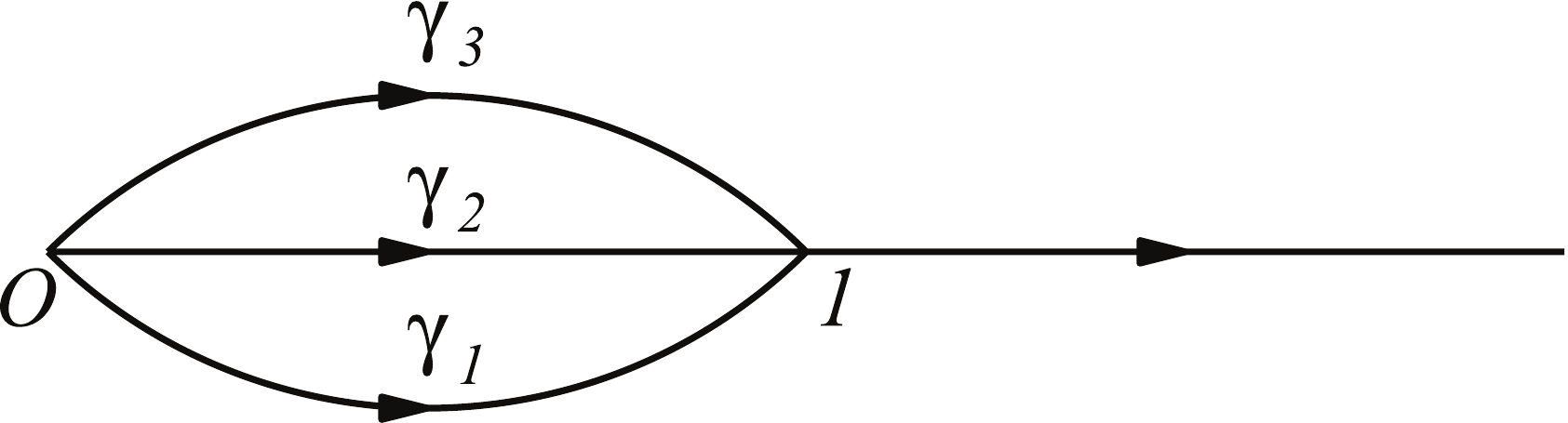}
   \end{center}
  \caption{\small{Contours  for the  RH problem for $S(z)$  in the $z$-plane.}}
 \label{contour-for-S}
    \end{figure}

Let us take a closer look at the function $\phi(z)$, defined in \eqref{phi-function} for $z\in \mathbb{C}\backslash [0, \infty)$.  We see that $\phi(x)>0$ for $x>1$, $\Re \phi(z)<0$ in the lens-shaped domains; cf. Figure \ref{contour-for-S}, and that $\phi_\pm (x)= \pm 2i \int^x_0\sqrt{\frac{1-s} s }ds$, purely imaginary, for $x\in (0,1)$.
Hence the RH problem for $T$ is oscillatory, in the sense that
the jump matrix in (\ref{T-jump-in-phi}) has  oscillating  diagonal
entries on the interval $(0,1)$. To remove the
oscillation, we introduce
 the second transformation $T\longrightarrow S$,  based on a factorization of
the oscillatory jump matrix and a deformation of contours.
We define
\begin{equation}\label{transformationT-S}
S(z)=\left \{
\begin{array}{ll}
  T(z), & \mbox{for $z$ outside the lens shaped region;}
  \\  [.4cm]
  T(z) \left( \begin{array}{cc}
                                 1 & 0 \\
                                  - z^{-\alpha}e^{2n\phi(z)} & 1 \\
                               \end{array}
                             \right) , & \mbox{for $z$ in the upper lens
                             region;}\\[.4cm]
T(z) \left( \begin{array}{cc}
                                 1 & 0 \\
                                   z^{-\alpha} e^{2n\phi(z)} & 1 \\
                               \end{array}
                             \right) , & \mbox{for  $z$  in  the  lower
                             lens
                             region, }
\end{array}\right .\end{equation}
where $\arg z\in (-\pi, \pi)$.
Then  $S$ solves the RH problem:
\begin{description}
  \item(S1)~~ $S(z)$ is analytic  in  $\mathbb{C}\backslash\Sigma_S$, where $\Sigma_S=\left\{\cup_{k=1}^3\gamma_k \right\} \cup(1,\infty)$,  illustrated in Figure \ref{contour-for-S};
  \item(S2)~~The jump conditions are
  \begin{equation}\label{S-jump}S_+(z)=S_-(z)  \left\{\begin{array}{ll}
                 \left( \begin{array}{cc}
                                 1 & 0 \\
                                  z^{-\alpha} e^{2n\phi(z)} & 1 \\
                               \end{array}
                             \right), &  z\in \gamma_1\cup \gamma_3,\\ [.4cm]
               \left( \begin{array}{cc}
                                0&   x^{\alpha} \\
                                   -x^{-\alpha} & 0 \\
                               \end{array}
                             \right), &  z=x\in \gamma_2, \\ [.4cm]
                  \left( \begin{array}{cc}
                                 1 &  z^{\alpha}e^{-2n \phi(z)} \\
                                  0  & 1 \\
                               \end{array}
                             \right), &   z\in (1,+\infty);
                \end{array}\right.\end{equation}
  \item(S3)~~ The asymptotic behavior at infinity is
  \begin{equation}\label{S-infty}S(z)= I+O( 1/ z),~~~\mbox{as}~~z\rightarrow
                             \infty  ;\end{equation}
 \item(S4)~~ The asymptotic behavior at the origin is sector-wise. As $z\rightarrow  0$,
 \begin{equation}\label{S-origin}S(z)=O(1)e^{-\frac {t_n}{8nz}\sigma_3}
 \left\{
 \begin{array}{ll}
   I,  & \mbox{outside~the~lens-shaped~regions}, \\[.4cm]
   \left( \begin{array}{cc}
                                 1 & 0 \\
                                  -z^{-\alpha} e^{2n\phi(z)} & 1 \\
                               \end{array}
                             \right), & \mbox{in~the~upper~lens~region},  \\[.4cm]
   \left( \begin{array}{cc}
                                 1 & 0 \\
                                  z^{-\alpha} e^{2n\phi(z)} & 1 \\
                               \end{array}
                             \right), & \mbox{in~the~lower~lens~region}.
 \end{array}
\right .
\end{equation}
\end{description}

\subsection{Global parametrix}\label{sec:3.4}

From (\ref{S-jump}), we see  that the jump matrix for $S$ is of the
form $J_{S, j}=I$,  plus an exponentially  small  term for
fixed $z\in \gamma_1\cup\gamma_3\cup (1, \infty)$. Neglecting the exponential small terms, we arrive at an approximating  RH problem for $N(z)$, as follows:
\begin{description}
\item(N1)~~  $N(z)$ is analytic  in  $\mathbb{C}\backslash
[0,1]$;
\item(N2)~~   \begin{equation}\label{N-jump} N_{+}(x)=N_{-}(x)\left(
       \begin{array}{cc}
       0 & x^{\alpha} \\
       -x^{-\alpha} & 0 \\
       \end{array}
       \right)~~~\mbox{for}~~x\in  (0,1);\end{equation}
\item(N3)~~    \begin{equation}\label{N-infty} N(z)= I+O( 1/ z) ,~~~\mbox{as}~~z\rightarrow\infty .\end{equation}
  \end{description}

A solution to the above RH problem can be constructed explicitly,
 \begin{equation}\label{N-solution} N(z)= D_{\infty}^{\sigma_3}M^{-1}    a (z)^{-\sigma_3}   MD(z)^{-\sigma_3}, \end{equation}
where $M=(I+i\sigma_1) /{\sqrt{2}}$, $a
(z)=\left(\frac{z-1} {z}\right)^{1/4}$ with $\arg z\in (-\pi, \pi)$ and $\arg (z-1)\in (-\pi, \pi)$, and the Szeg\"{o} function
 \begin{equation*} D(z)=\left(\frac{z}{\varphi(2z-1)}\right)^{\alpha/2},~~\varphi(z)=z+\sqrt{z^2-1}, \end{equation*}
the branch is chosen such that $\varphi(z) \sim  2z$ as $z\rightarrow\infty$, and  $D_{\infty}=2^{-\alpha}$.

The jump matrices of $SN^{-1}$ are not uniformly close to the unit matrix near the end-points $0$ and $1$, thus local parametrices have to be constructed  in
neighborhoods of the end-points.

\subsection{Local parametrix $P^{(1)}(z)$ at $z=1$ }\label{sec:3.5}
The local parametrix at the right  end-point $z=1$ is the same
as that of the Hermite polynomials or the Laguerre polynomials at the soft edge. More precisely, the parametrix is to be constructed in
$U(1,r)=\{z~|\;|z-1|<r\}$,   $r$ being a fixed positive number, such that
\begin{description}
  \item(a)~~ $P^{(1)}(z)$ is analytic in $U(1,r) \backslash \Sigma_{S}$, see Figure \ref{contour-for-S} for the contours $\Sigma_{S}$;
  \item(b)~~ In  $U(1,r)$, $P^{(1)}(z)$ satisfies the same jump conditions as $S(z)$ does; cf. (\ref{S-jump});
  \item(c)~~  $P^{(1)}(z)$ fulfils the following  matching condition
  on $\partial U(1,r)$:
\begin{equation}\label{matchingP1-N}
P^{(1)}(z)N^{-1}(z)=I+ O\left ( 1 /n\right ).
 \end{equation}
\end{description}
The parametrix can be constructed, out   of the Airy function
and its derivative, as  in Section \ref{sec:6.1} below, and in \cite[(3.74)]{vanlessen}; see also  \cite{deift,dkmv2,qw}.

\subsection{Local parametrix $P^{(0)}(z)$ at the origin}
In this subsection, we focus on the construction of the
parametrix at  $z=0$. The parametrix, to be
constructed  in the neighborhood  $U(0,r)=\{z~|\;|z|<r\}$  for sufficiently small $r$, solves a RH problem as
follows:
\begin{description}
  \item(a)~~ $P^{(0)}(z)$ is analytic in $U(0,r) \backslash  \Sigma_{S}$;
  \item(b)~~ In  $U(0,r)$, $P^{(0)}(z)$ satisfies the same jump conditions as $S(z)$ does; cf. (\ref{S-jump});
  \item(c)~~  $P^{(0)}(z)$ fulfils the following  matching condition
   on  $\partial U(0,r)=\{~z\; |\; |z|=r\}$:
\begin{equation}\label{matchingP0-N}
P^{(0)}(z)N^{-1}(z)=I+ O\left ( n^{-1/3}\right )~~\mbox{as}~n \to \infty;
 \end{equation}
 \item(d)~~  The behavior at the center  $z=0$ is the same as  that of  $S(z)$, as described in (\ref{S-origin}).
 \end{description}

Now we apply a  transformation to convert all  the jumps of  the RH problem for $P^{(0)}(z)$  to constant jumps by defining
\begin{equation}\label{TransfP0-hat-P0}P^{(0)}(z)=\hat{P}^{(0)}(z)(-z)^{-\frac \alpha2\sigma_3}e^{n\phi(z)\sigma_3}, ~~z\in U(0,r) \backslash  \Sigma_{S},\end{equation}where $\arg(-z)\in (-\pi, \pi)$.
It is readily seen  that $\hat{P}^{(0)}$ solves the RH problem

\begin{description}
  \item(a)~~ $\hat{P}^{(0)}(z)$ is analytic in $U(0,r) \backslash  \Sigma_{S}$;
  \item(b)~~ In  $U(0,r)$, $\hat{P}^{(0)}(z)$ satisfies the   jump conditions
   \begin{equation}\label{P0-hat-jump}
 \hat{P}^{(0)}_+(z)=\hat{P}^{(0)}_-(z)
 \left\{\begin{array}{ll}
          \left(
                               \begin{array}{cc}
                                 1 & 0 \\
                               e^{-\pi i\alpha}& 1 \\
                                 \end{array}
                             \right), & z\in \gamma_3 \cap  U(0,r), \\[.4cm]
          \left(
                               \begin{array}{cc}
                                0 &1\\
                               -1&0 \\
                                 \end{array}
                             \right),  &  z\in (0,r), \\[.4cm]
           \left(
                               \begin{array}{cc}
                                 1 &0 \\
                                  e^{\pi i\alpha} &1 \\
                                 \end{array}
                             \right), & z\in \gamma_1 \cap U(0,r);
        \end{array}\right . \end{equation}
   \item(c)~~  The behavior at the center  $z=0$ is, as $z\to 0$,
 \begin{equation}\label{P0-hat-origin}\hat{P}^{(0)}(z)=O(1)e^{-\frac {t_n}{8nz}\sigma_3}
  (-z)^{\frac \alpha 2 \sigma_3} e^{-n\phi(z)\sigma_3}
 \left\{
 \begin{array}{ll}
   I,  & \mbox{outside~the~lens}, \\[.4cm]
   \left( \begin{array}{cc}
                                 1 & 0 \\
                                  -e^{-i\pi\alpha} & 1 \\
                               \end{array}
                             \right), & \mbox{upper~lens},  \\[.4cm]
   \left( \begin{array}{cc}
                                 1 & 0 \\
                                  e^{i\pi\alpha} & 1 \\
                               \end{array}
                             \right), & \mbox{lower~lens}.
 \end{array}
\right .
 \end{equation}
 \end{description}

We are now in a position to bring together $\hat{P}^{(0)}$ and the    model problem \eqref{Psi-jump}-\eqref{Psi-origin}, treated  in the previous section. To this aim, we note first that
\begin{equation}\label{conformal-mapping}\zeta=n^2\phi^2(z) \end{equation} is a conformal mapping in the $z$-neighborhood $U(0,r)$, for $r$ sufficiently small, such that $\zeta   \approx   -16n^2 z$ for small $z$; cf. \eqref{phi-function}. Specifying the parts of the contours $\gamma_1-\gamma_3$ within $U(0,r)$ so that they map respectively  to the rays $\Sigma_1-\Sigma_3$; see Figure \ref{contour-for-model} and Figure \ref{contour-for-S} for the contours.
Attention should be paid to the orientation of the contours, which has been reversed after the conformal mapping.

Based on these, we  seek a solution $\hat{P}^{(0)}(z)$ of the following form
 \begin{equation}\label{P0-hat-solution}\hat{P}^{(0)}(z)=E(z)\Psi(n^2\phi^2, 2nt_n)e^{-\frac \pi 2 i\sigma_3},\end{equation}
 where $E(z)$ is an analytic matrix in  $U(0,r)$,  taken  to meet the matching  condition (\ref{matchingP0-N}), and  the  factor  $e^{-\frac \pi 2 i\sigma_3}$
 is appended in accordance with the now reversed  orientation of $\Sigma_1-\Sigma_3$; cf.  \eqref{phi-function} and the fact that  the conformal mapping $\zeta=n^2\phi^2\approx -16n^2z$.

 From \eqref{P0-hat-solution} and \eqref{Psi-jump}-\eqref{Psi-origin}, it is readily verified that the jumps \eqref{P0-hat-jump} and the behavior at the origin \eqref{P0-hat-origin} is well fulfilled. What is more, the analytic factor can be determined by  the matching condition (\ref{matchingP0-N}) and the asymptotic behavior of $\Psi(\zeta)$ for $\zeta\to\infty$. Indeed,  we can take
\begin{equation}\label{P0-hat-analytic-factor}E(z)=N(z)e^{\frac \pi 2 i\sigma_3} (-z)^{\frac \alpha 2\sigma_3}\frac{I-i\sigma_1}{\sqrt{2}}
\left\{n^2\phi^2(z)\right\}^{\frac 1 4\sigma_3},\end{equation}
where $\arg(-z)\in (-\pi, \pi)$, and  $\arg  \left\{ n^2\phi^2(z)\right\}   \in (-\pi, \pi)$.
The matching condition \eqref{matchingP0-N} now follows from  \eqref{N-solution}, \eqref{P0-hat-solution}-\eqref{P0-hat-analytic-factor} and \eqref{Psi-infty}; see also Remark \ref{rem-error-R} below.

At last, we show that, so-defined $E(z)$ is an analytic function in $U(0, r)$. Since $E(z)$ is analytic in $U(0,r)\backslash \gamma_2$, it suffices to show that
\begin{equation*}  E_+(x)=E_-(x)~~\mbox{for}~x\in \gamma_2,\end{equation*} and that $E(z)$ possesses at most  a weak singularity at  $z=0$, both can be verified straightforward from \eqref{P0-hat-analytic-factor} and \eqref{N-solution}. Here use may also be made of the facts that $\arg(-z)=\mp\pi$ respectively on the positive and negative side of $\gamma_2$, and also that $\arg (n^2\phi^2(z))= \mp\pi$, from above or below $\gamma_2$.

\subsection{The final transformation $S\rightarrow R$}\label{sec:3.7}

\begin{figure}[t]
 \begin{center}
   \includegraphics[width=9cm]{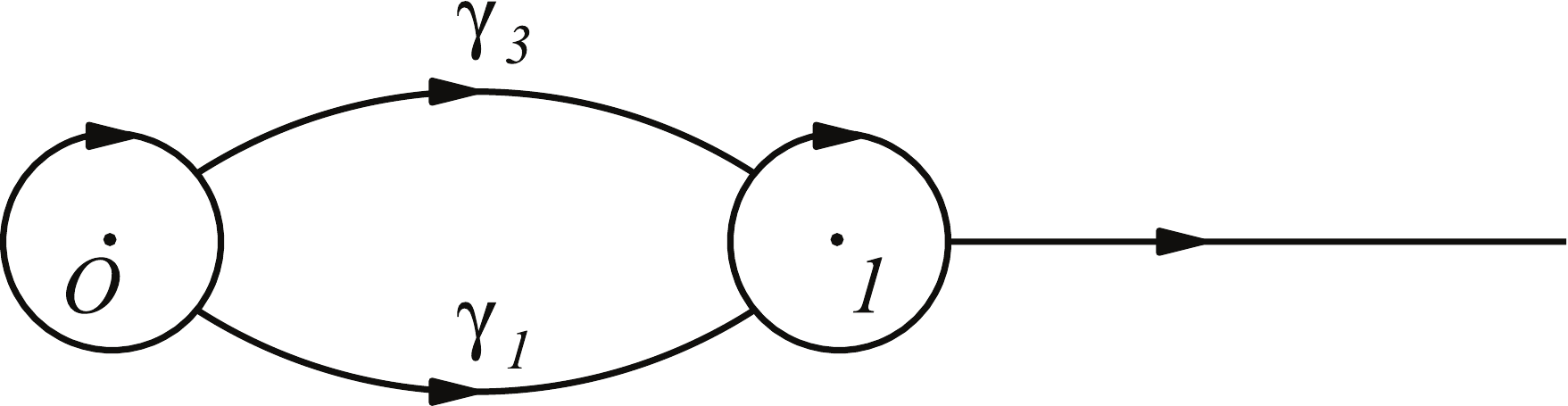} \end{center}
  \caption{\small{Contours  for the  RH problem for $R(z)$  in the $z$-plane.}}
 \label{contour-for-R}
    \end{figure}

Now we bring in the final transformation by defining
\begin{equation}\label{transformationS-R}
R(z)=\left\{ \begin{array}{ll}
                S(z)N^{-1}(z), & z\in \mathbb{C}\backslash \left \{ U(0,r)\cup U(1,r)\cup \Sigma_S \right \};\\[.1cm]
               S(z) (P^{(0)})^{-1}(z), & z\in   U(0,r)\backslash \Sigma_{S} ;  \\[.1cm]
               S(z)  (P^{(1)})^{-1}(z), & z\in   U(1,r)\backslash
               \Sigma_{S} .
             \end{array}\right .
\end{equation}
Then, $R(z)$ solves the following RH problem:
\begin{description}
  \item(R1)~~ $R(z)$ is analytic in $\mathbb{C}  \backslash \Sigma_R$ (see Figure  \ref{contour-for-R} for the contours);
  \item(R2)~~ $R(z)$ satisfies the  jump conditions
  \begin{equation}\label{R-jump}
R_+(z)=R_-(z)J_{R}(z), ~~z\in\Sigma_R,\end{equation}where
$$J_R(z)=\left\{ \begin{array}{ll}
                    P^{(0)}(z)N^{-1}(z), ~ &   z\in\partial U(0,r),\\[.1cm]
                     P^{(1)}(z)N^{-1}(z),&  z\in\partial
                    U(1,r),\\[.1cm]
N(z)J_{S,j}(z)N^{-1}(z), ~& \Sigma_R\setminus \partial
                   ( U(0,r)\cup  U(1,r));
                 \end{array}\right .$$
  \item(R3)~~  $R(z)$ demonstrates the following behavior at infinity:
  \begin{equation}\label{R-infty}
  R(z)= I+ O\left({1}/{z} \right),~~\mbox{as}~ z\rightarrow\infty .
  \end{equation}
\end{description}

It follows from the matching condition (\ref{matchingP0-N}) of the local parametrices and the definition of $\phi$ that
\begin{equation}\label{R-jump-approx}J_R(z)=\left\{ \begin{array}{ll}
                     I+O\left (n^{-1/3}\right ),&  z\in\partial
                    U(0,r)\cup  U(1,r),\\[.1cm]
I+O(e^{-cn}), ~& z\in  \Sigma_R\setminus \partial
                   ( U(0,r)\cup  U(1,r)),
                 \end{array}\right .
\end{equation}
where $c$ is a positive constant, and the error term is uniform
for $z$ on the corresponding contours. Hence we have
\begin{equation}\label{R-jump-estimate}\|J_R(z)-I\|_{L^2\cap L^{\infty}(\Sigma_R)}=O(n^{-1/3}).
\end{equation}
Then, applying  the now standard  procedure of norm
estimation  of Cauchy operator and using the technique of deformation of contours (cf. \cite{deift,dkmv2}), it
follows from   (\ref{R-jump-estimate}) that
 \begin{equation}\label{R-approx}R(z)=I+O(n^{-1/3}),
\end{equation} uniformly for $z$ in the whole complex plane.

This completes the nonlinear steepest descent analysis.

{\rem{\label{rem-error-R}
Indeed, for $s=2nt$ in compact subsets of $(0, \infty)$, the error in  \eqref{matchingP0-N}   and in \eqref{R-jump-approx}  can be made uniformly  $O(1/n)$.   For $t$ in larger range, $t\in (0, d]$ for fixed $d$, detailed calculation shows that the error term takes the weaker form  $O( {n^{-1/3}})$,  with the uniformity preserved, as will be confirmed by Theorem \ref{thm-transition-to-Bessel} and Theorem \ref{thm-transition-to-Airy}, see also   \eqref{Psi-in-Bessel-approx}  and  (\ref{Psi-representation-outside}) below.}}

\section{Proof of   Theorem \ref{thm-limit-kernel}: Limiting kernel at the edge}\label{sec:4}

In terms of the matrix-valued function $Y(z)$ defined in \eqref{Y-solution},  the kernel $K_n(x,y)$ in \eqref{kernel-CD-formula}
 can be written as
\begin{equation*}
K_n(x,y)=\frac{\sqrt{w(x)w(y)}}{2\pi i(x-y)}\left\{
Y_+^{-1}(y)Y_+(x)\right \}_{21},
\end{equation*}where $x$ and $y$ belong to the support of the equilibrium measure, which in this case is $[0, 4n]$. So it is natural to introduce a re-scaling of the variable to fix the support to $[0,1]$, and to consider the  re-scaled kernel $4nK_n(4nx, 4ny)$, such  that
 \begin{equation}\label{kernel-in-Y}
\tilde K_n(x,y):=  4nK_n(4nx, 4ny)  =\frac{\sqrt{w(4nx)w(4ny)}}{2\pi i (x-y)}\left\{
Y_+^{-1}(4ny)Y_+(4nx)\right \}_{21},~~x,y\in(0,1).
\end{equation}
We note that such a re-scaling amounts to considering a so-called varying weight, in this case, the weight is an re-scaled version of \eqref{weight-of-the-paper}, namely,
\begin{equation*}
\tilde w(x;n)=4n\,  w(4nx; t_n)=  (4n )^{\alpha+1}  x^\alpha e^{-4nx-t_n/(4nx)}~~\mbox{for}~x\in (0, \infty).
\end{equation*}
However, in the asymptotic analysis  conducted in the previous section, we have chosen to analyze the original $w(x; t)$ in \eqref{weight-of-the-paper}.

Represented in \eqref{kernel-in-Y}, the asymptotics of the kernel, or, the large-$n$ limit of it, can be derived   from the  Riemann-Hilbert analysis.
We note that the large-$n$ limit   for $\tilde K_n(x,y)$ is expect to be the sine kernel for $x, y$ in the interior of the equilibrium measure, namely, $x, y\in (0, 1)$, and at the soft edge $x=1$, the kernel is expected to be approximated by the Airy kernel. The reader is referred to \cite{vanlessen} for a detailed analysis.  In the present paper, we focus on the edge behavior at the end-point $z=0$, where there is an essential singularity of the weight function $w(x; t)$. Attention will be paid to the dependence on $t=t_n$ of the statistic quantities.

Tracing back the transformations $R\to S\to T\to Y$, and combining    \eqref{TrsnaformationY-T}, \eqref{transformationT-S} and \eqref{transformationS-R} with \eqref{TransfP0-hat-P0} and \eqref{P0-hat-solution},    we have
\begin{equation}\label{Y-trace-back}Y_+(4nx)=c_n^{\sigma_3} R(x) E(x) \Psi_- (f_n(x), s)  e^{\frac {i\pi} 2   (\alpha-1 )\sigma_3}\left(
                                                                                                                \begin{array}{cc}
                                                                                                                  1 & 0 \\
                                                                                                                 1 & 1 \\
                                                                                                                \end{array}
                                                                                                              \right)\left [w(4nx)\right ]^{-\frac 1 2\sigma_3},\end{equation}
for $0<x<r$, where
\begin{equation*}  f_n(x) =n^2\phi^2(x), \quad s=2nt_n, \quad  c_n= (-1)^n (4n)^{n+\frac \alpha 2} e^{\frac 1 2 {nl}}, \end{equation*} and use has been made of the fact  that $e^{n\pi i\sigma_3}=(-1)^n I$,
\begin{equation*}  g_+(x)+\phi_+(x) -\frac 1 2l =2x+\pi i~~\mbox{for}~x\in (0, 1), \end{equation*}as can be seen from Section \ref{sec:3.2}, and that the boundary value on the positive side $P^{(0)}_+$ corresponds to the value $\Psi_-$ on the negative side, as can be seen from the correspondence $\zeta \approx -16n^2 z$, and the orientation of   $\gamma_2$ and $\Sigma_2$; cf. Figures  \ref{contour-for-model} and \ref{contour-for-S}.
   Substituting \eqref{Y-trace-back} into \eqref{kernel-in-Y},  we have
\begin{equation}\label{kernel-representation}
\tilde K_n(x,y)= \frac{
\left(
    - \psi_2\left (f_n(y)\right ), \psi_1\left (f_n(y)\right )\right )
E^{-1}(y)R^{-1}(y)R(x)E(x)
\left(
     \psi_1\left (f_n(x)\right ), \psi_2\left (f_n(x)\right)
\right)^T}{2\pi i(x-y)},
 \end{equation}
 where $X^T$ stands for the transpose matrix of $X$, and
 \begin{equation}\label{psi1-psi2-def}
  \left(
                                \begin{array}{c}
                                  \psi_1(\zeta) \\
                                  \psi_2(\zeta) \\
                                \end{array}
                              \right) =
 \left(
                                \begin{array}{c}
                                  \psi_1(\zeta,s) \\
                                  \psi_2(\zeta,s) \\
                                \end{array}
                              \right)
 =\left (\Psi\right )_- \left( \zeta,s\right )         \left(
                                                           \begin{array}{c}
                                                           e^{\frac{\pi}2 i (\alpha-1)}  \\ e^{-\frac{\pi}2 i (\alpha-1)}
                                                           \end{array}
                                                         \right)~~ \mbox{for}~\zeta\in (-\infty, 0). \end{equation}

Now let  $x=\frac {u}{16n^2}$ and $ y=\frac {v}{16n^2}$, where both $u$ and $v$ are positive and of the size $O(1)$. In view of \eqref{phi-function}, we obtain  \begin{equation*} f_n(z)=  n^2\phi^2(z)=n^2  \left [ -16z+O(z^2)\right ] \end{equation*}  for small $z$. In particular, we have
\begin{equation*} f_n(x)=-u\left [ 1+O\left (\frac 1{n^2}\right )\right ],~~\mbox{and}~f_n(y)=-v\left [1+O\left (\frac 1{n^2}\right )\right ].\end{equation*}

The analyticity  of $E(z)$ in $U(1, r)$  implies that both $E(z)$ and $E^{-1}(z)$ are bounded in a neighborhood of the origin, and
\begin{equation}\label{E-inverse-E}E^{-1}(y)E(x)=I+E^{-1}(y)(E(x)-E(y))=I+O(x-y)=I+(u-v) O\left (n^{-2}\right ) \end{equation}
for  bounded $u$, $v$.
Similarly,
since $R(z)$ is a matrix function analytic in $U(0, r)$, we have
\begin{equation}\label{R-inverse-R}R^{-1}(y)R(x)=I+(u-v) O\left (n^{-2}\right ).\end{equation}
Here again,  the error term  is uniform for    $u$ and $v$ lying in compact subsets of $(0, \infty)$. Also we have
\begin{equation}\label{psi-k-approx}\psi_k(f_n(x), s)=\psi_k(-u, s) +O\left (n^{-2}\right ) \end{equation}for $k=1,2$, and the error bound is uniform for both  $u$  and  $s$  in compact subsets of $(0, \infty)$.

Thus, substituting \eqref{E-inverse-E}, \eqref{R-inverse-R} and \eqref{psi-k-approx} into   \eqref{kernel-representation},  we have
\begin{equation}\label{kernel-approx-almost-final}
\frac {1}{4n} K_n\left (\frac {u}{4n} ,\frac {v}{4n}\right )= \frac{  \psi_1(f_n(y),s)\psi_2(f_n(x),s)   -  \psi_1(f_n(x),s)\psi_2(f_n(y),s)}{2\pi i(u-v)} +O\left (\frac 1 {n^2}\right)
 \end{equation}for large $n$.

We introduce   an auxiliary function with two variables
\begin{equation*} H(\xi, \eta)=  \frac{  \psi_1(\eta ,s)\psi_2(\xi, s)   -  \psi_1(\xi, s)\psi_2(\eta ,s)}{2\pi i( \eta-\xi)}, ~~\xi,\eta\in (-\infty, 0).  \end{equation*}
For $s\in (0, \infty)$ fixed,   $\psi_k(\eta ,s)$, $k=1,2$ can be extended to an analytic function. It is thus easily seen that $H(\xi, \eta)$ is $C^\infty$ in $(-\infty, 0)\times (-\infty, 0)$, noting that there is no singularity on the ray $\xi=\eta$. Therefore we have
\begin{equation*} H(\xi , \eta )= H(\xi_0, \eta_0)+ O(| \xi- \xi_0|+| \eta-\eta_0 |) .  \end{equation*}
Substituting  $\xi=f_n(x)=f_n\left (\frac u{16n^2}\right )$, $\eta=f_n(y)=f_n\left (\frac v{16n^2}\right )$, $\xi_0= -u$ and $\eta_0=-v$ into the above equation, we obtain
\begin{equation}\label{kernel-approx-final}
\frac {1}{4n} K_n\left (\frac {u}{4n} ,\frac {v}{4n}\right )= \frac{  \psi_1(-v,s)\psi_2(-u,s)   -  \psi_1(-u,s)\psi_2(-v,s)}{2\pi i(u-v)}+O\left (\frac 1 {n^2}\right)
 \end{equation}for large $n$,
where the error term $O\left (n^{-2}\right)$ is uniform for $u$ and $v$  in  compact subsets of  $(0,\infty)$ and uniformly for $s\in (0,\infty)$.   In deriving the last formula, use has also been made of the fact that
\begin{equation*} \frac { \xi -\eta} {v-u} =1+O\left ( \frac 1 {n^2}\right ),  \end{equation*}uniformly for $u, v$ belong to compact subsets of $(0, \infty)$ and for large $n$.

 This is exactly \eqref{kernel-approx-final-introduction}. It thus completes the proof of   Theorem \ref{thm-limit-kernel}.

\section{Proof of Theorem \ref{thm-transition-to-Bessel}: Transition to the Bessel kernel as $s\rightarrow 0^+$}\label{sec:5}

In this section, we study the    asymptotics of the model RH problem for $\Psi(\zeta,s)$ for small positive  parameter $s$.
Then we    apply the results to    reduce the  $\Psi$-kernel in \eqref{kernel-approx-final-introduction}  to a Bessel kernel, and to obtain  initial conditions for the equations of $r$, $t$ and $q$, derived in \eqref{equation-sets} and \eqref{r-eqn}. A similar discussion can be found in \cite{xz2013b}.

\subsection{Nonlinear steepest descend analysis of  the model RH  problem as $s\rightarrow0^+$}
If $s=0$, in the model RH problem  for $\Psi(\zeta,s)$; cf.  (\ref{Psi-jump})-(\ref{Psi-origin}),  the essential singularity at the origin vanishes. So, to consider the approximation of  $\Psi(\zeta,s)$ for small $s$, it is natural to  ignore temporarily  the exponential term $e^{\frac {s}\zeta}$ in \eqref{Psi-origin}, and to consider   a limiting  RH problem  $\Psi_0(\zeta)$.

\begin{figure}[h]
 \begin{center}
   \includegraphics[width=8cm]{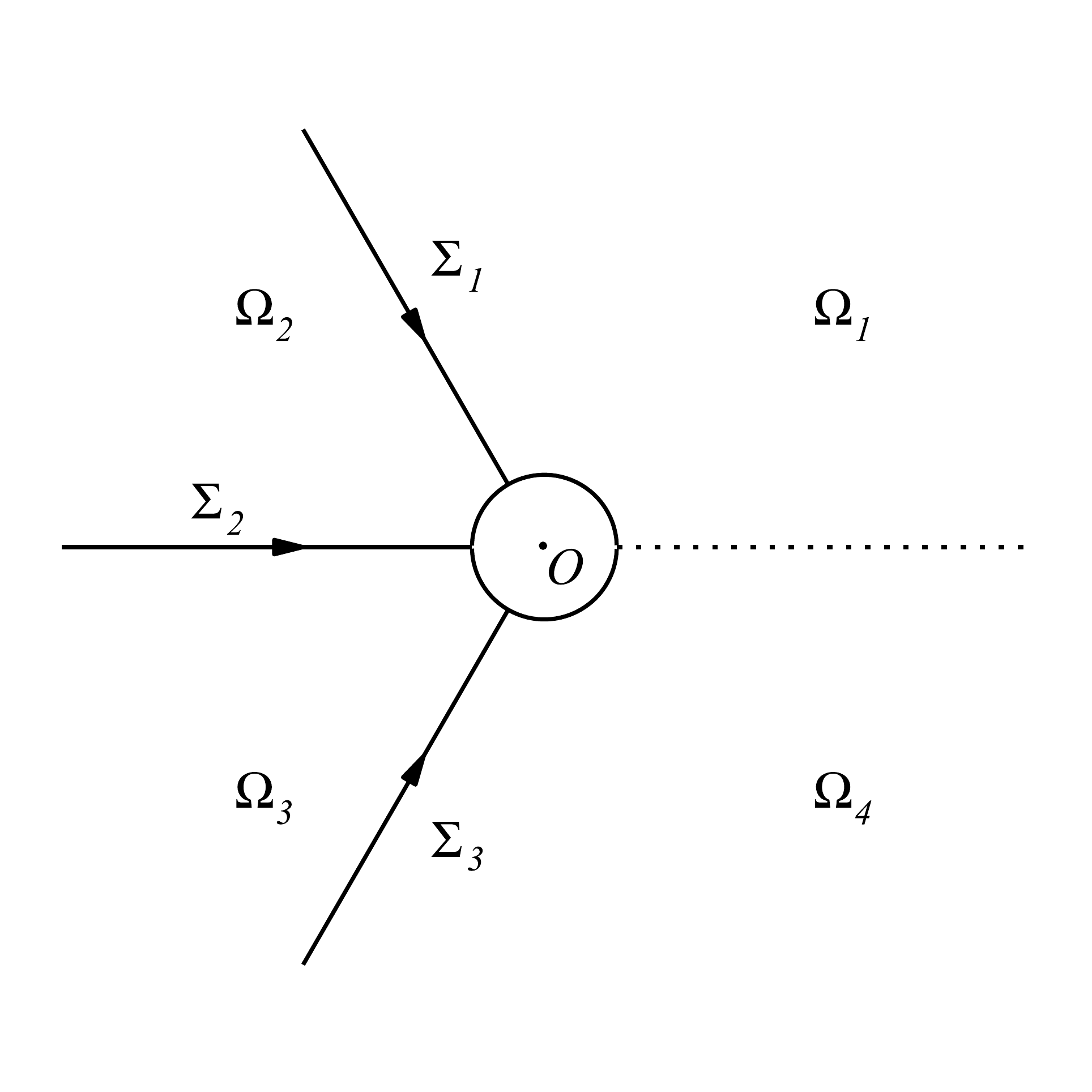} \end{center}
  \caption{\small{Contours  for the  $\Psi$ to Bessel transition in the $\zeta$-plane.}}
 \label{contour-for-pro1}
    \end{figure}

 \begin{description}
  \item(a)~~  $\Psi_0(\zeta)$ is analytic in
  $\mathbb{C}\backslash\cup^3_{j=1}\Sigma_j$ (see Figure  \ref{contour-for-pro1});

  \item(b)~~  $\Psi_0(\zeta)$  satisfies the jump condition
\begin{equation}\label{Psi-0-jump}
\left (\Psi_0\right )_+ (\zeta)=\left (\Psi_0\right )_- (\zeta)
\left\{ \begin{array}{ll}
          \left(
                               \begin{array}{cc}
                                 1 & 0 \\
                               e^{\pi i\alpha}& 1 \\
                                 \end{array}
                             \right),       &  \zeta \in \Sigma_1 ,\\[.4cm]
           \left(
                               \begin{array}{cc}
                                0 &1\\
                               -1&0 \\
                                 \end{array}
                             \right), &   \zeta \in \Sigma_2 ,\\[.4cm]
                \left(
                               \begin{array}{cc}
                                 1 &0 \\
                                  e^{-\pi i\alpha} &1 \\
                                 \end{array}
                             \right), &         \zeta \in \Sigma_3;     \end{array} \right .
    \end{equation}

\item(c)~~  The asymptotic behavior of $\Psi_0(\zeta)$  at infinity
  is
  \begin{equation}\label{Psi-0-infty}
\Psi_0(\zeta)=
    \left (I+O\left (\frac 1{ \zeta}\right )\right)\zeta^{-\frac{1}{4}\sigma_3}\frac{I+i\sigma_1}{\sqrt{2}}
 e^{\sqrt{\zeta}\sigma_3}, ~~\arg\zeta \in (-\pi, \pi),~~\zeta\to \infty.
  \end{equation}
  \end{description}

\medskip

 A function  $\Phi(\zeta)$, satisfying the jump \eqref{Psi-0-jump} and the behavior \eqref{Psi-0-infty} at infinity,     can be constructed in terms of the modified Bessel functions as (cf. \cite{kmvv})
 \begin{equation}\label{Phi-Bessel}
 \Phi(\zeta)=M_1 \pi ^{\frac 1 2\sigma_3}
 \left\{
 \begin{array}{ll}
   \left(
                               \begin{array}{cc}
                                 I_{\alpha}(\sqrt{\zeta}) &\frac{i}{\pi} K_{\alpha}(\sqrt{\zeta})  \\
                                 \pi i\sqrt{\zeta}    I'_{\alpha}(\sqrt{\zeta})& -\sqrt{\zeta}    K'_{\alpha}(\sqrt{\zeta}) \\
                                 \end{array}
                             \right),  &    \zeta \in \Omega_1\cup \Omega_4, \\[0.5cm]
  \left(
                               \begin{array}{cc}
                                 I_{\alpha}(\sqrt{\zeta}) &\frac{i}{\pi} K_{\alpha}(\sqrt{\zeta})  \\
                                 \pi i\sqrt{\zeta}    I'_{\alpha}(2\sqrt{\zeta})& -\sqrt{\zeta}    K'_{\alpha}(\sqrt{\zeta}) \\
                                 \end{array}
                             \right)  \left(
                               \begin{array}{cc}
                                 1 & 0 \\
                                 -e^{\pi i\alpha} & 1 \\
                                 \end{array}
                             \right),    &    \zeta\in\Omega_2,  \\[0.5cm]
   \left(
                               \begin{array}{cc}
                                 I_{\alpha}(\sqrt{\zeta}) &\frac{i}{\pi} K_{\alpha}(\sqrt{\zeta})  \\
                                 \pi i\sqrt{\zeta}    I'_{\alpha}(\sqrt{\zeta})& -\sqrt{\zeta}    K'_{\alpha}(\sqrt{\zeta}) \\
                                 \end{array}
                             \right)\left(
                               \begin{array}{cc}
                                 1 & 0 \\
                                 e^{-\pi i\alpha} & 1 \\
                                 \end{array}
                             \right), &  \zeta\in\Omega_3,
 \end{array}
 \right . \end{equation}where $\arg \zeta\in (-\pi, \pi)$, and  $M_1=\left(
                                     \begin{array}{cc}
                                       1 & 0 \\
                                       \frac i 8(4\alpha^2+3) & 1\\
                                     \end{array}
                                   \right)
 $, the regions are illustrated in Figure \ref{contour-for-pro1}.
Indeed, referring to \cite[(10.27.6) and  (10.27.9)]{nist}, it is easily seen that  \eqref{Psi-0-jump} is satisfied by $\Phi(\zeta)$. By expanding the modified Bessel functions   for large $\zeta$ (see \cite[\S10.27 and \S10.40]{nist}),  we have a more  precise version of \eqref{Psi-0-infty}
 \begin{equation}\label{Phi-Bessel-infty}\Phi(\zeta)=\left[I+\frac {4\alpha^2-1}{128\zeta}\left(
                                           \begin{array}{cc}
                                             4\alpha^2-9 & 16i \\
                                            \frac {i}{12} (4\alpha^2-9)(4\alpha^2-13) & 9-4\alpha^2 \\
                                           \end{array}
                                         \right) +O\left (\zeta^{-\frac 32}\right )\right ]\zeta^{-\frac{1}{4}\sigma_3}M
 e^{\sqrt{\zeta}\sigma_3},
 \end{equation}where $M= {(I+i\sigma_1)}/{\sqrt{2}}$.

 However, at the origin, the asymptotic behavior of $\Phi(\zeta)$ is significantly different from that of  $\Psi(\zeta,s)$, as long as $s\not=0$.  When $\alpha\not\in \mathbb{Z}$, the behavior of the Bessel model RH problem at $\zeta=0$ takes the form
 \begin{equation}\label{Phi-Bessel-origin-equal}
 \Phi(\zeta)=\hat{\Phi}(\zeta)\zeta^{\frac{\alpha\sigma_3}{2}}\left(
                                                          \begin{array}{cc}
                                                            1 &\frac{1}{2i\sin (\alpha\pi)}  \\
                                                            0 & 1 \\
                                                          \end{array}
                                                        \right) J,~~\arg\zeta\in (-\pi, \pi),~~\zeta\rightarrow 0,
\end{equation}
where
 \begin{equation*}J=\left\{\begin{array}{lr}
                     I & \zeta \in  \Omega_1\cup\Omega_4, \\[.3cm]
                    \left(
                               \begin{array}{cc}
                                 1 & 0 \\
                                 -e^{\pi i\alpha} & 1 \\
                                 \end{array}
                             \right),  &\zeta \in \Omega_2,  \\[.3cm]
                     \left(
                               \begin{array}{cc}
                                 1 & 0 \\
                                 e^{-\pi i\alpha} & 1 \\
                                 \end{array}
                             \right), & \zeta \in \Omega_3,
                   \end{array}  \right.
 \end{equation*}   and
 \begin{equation*}\hat{\Phi}(\zeta)=
\left(
  \begin{array}{cc}
   \zeta^{-\alpha/2}I_{\alpha}( \sqrt\zeta) & \frac {i}{2\sin(\alpha\pi)} \zeta^{\alpha/2}I_{-\alpha}( \sqrt\zeta) \\[.2cm]
  \pi i   \zeta ^{(1-\alpha)/2}I'_{\alpha}( \sqrt\zeta) & \frac {-\pi}{\sin(\alpha\pi)} \zeta^{(1+\alpha)/2}I'_{-\alpha}( \sqrt\zeta) \\
  \end{array}
\right) \end{equation*}  is an entire matrix function, as can be seen from   the convergent series expansion of the modified Bessel functions
\begin{equation*} I_{\alpha}(z)=\left (\frac z 2\right )^{\alpha}\sum_{k=0}^{\infty}\frac{(z^2/4)^k}{k!\Gamma(\alpha+k+1)},\quad \mbox{and}~K_{\alpha}(z)=\frac{\pi}{2}\frac{I_{-\alpha}(z)-I_{\alpha}(z)}{\sin (\alpha\pi)};
\end{equation*}cf. \cite[(10.25.2) and (10.27.4)]{nist}.

It is clear that $\Psi(\zeta, s)$ is not approximated by $\Phi(\zeta)$ near the origin for $s>0$.    So we need to construct a local
 parametrix $F(\zeta)$,  defined in $U(0,\epsilon)$ for a small $\epsilon$, and matches $\Phi(\zeta)$ on $|\zeta|=\epsilon$. More precisely,      $F(\zeta)$   is supposed to  solve  the following RH problem:
\begin{description}
  \item(a)~~  $F(\zeta)$ is analytic in
  $U(0,\epsilon)\backslash\cup^3_{j=1}\Sigma_j$  (see Figure \ref{contour-for-model} for $\Sigma_j$, $j=1-3$);
\item(b)~~  $F(\zeta)$  satisfies the same jump condition as $\Psi(\zeta,s)$, that is,
\begin{equation}\label{F-jump}
F_+(\zeta)=F_-(\zeta)
 \left\{
 \begin{array}{ll}
    \left(
                               \begin{array}{cc}
                                 1 & 0 \\
                               e^{\pi i\alpha}& 1 \\
                                 \end{array}
                             \right), &  \zeta \in \Sigma_1\cap U(0,\epsilon), \\[.4cm]
    \left(
                               \begin{array}{cc}
                                0 &1\\
                               -1&0 \\
                                 \end{array}
                             \right),  &  \zeta \in \Sigma_2\cap U(0,\epsilon), \\[.4cm]
    \left(
                               \begin{array}{cc}
                                 1 &0 \\
                                  e^{-\pi i\alpha} &1 \\
                                 \end{array}
                             \right), &   \zeta \in \Sigma_3\cap U(0,\epsilon);
 \end{array}  \right .  \end{equation}
\item(c)~~ On the circular boundary  $|\zeta|=\epsilon$,   a matching condition is fulfilled, such that
\begin{equation}\label{F-matching}F(\zeta)=\left [I+O(s)+O\left (s^{\alpha+1}\right )\right ]\Phi(\zeta)~~\mbox{as}~s\to 0^+; \end{equation}
\item(d)~~ The asymptotic behavior of $F(\zeta)$  at the origin is the same as that of $\Psi(\zeta,s)$ in \eqref{Psi-origin}, namely, \begin{equation}\label{F-origin}F(\zeta)=O(1)\zeta^{\frac \alpha 2\sigma_3}e^{\frac s \zeta\sigma_3}J,\end{equation}
  where $J$ is the constant factor introduced in \eqref{Phi-Bessel-origin-equal}, defined sector-wise.
 \end{description}

 We seek a solution  $F(\zeta)$ of the form
 \begin{equation}\label{F-solution}
 F(\zeta)=\hat\Phi(\zeta)  \left(
                               \begin{array}{cc}
                                 1 & f(\zeta ) \\
                                 0 & 1\\
                               \end{array}
                             \right)
 \zeta^{\frac{\alpha\sigma_3}{2}}e^{\frac s \zeta\sigma_3}J,
\end{equation}
where $\tilde\Phi(\zeta)$ is the same entire function as in \eqref{Phi-Bessel-origin-equal}, and the scalar function $f(\zeta)$, to be determined,  is analytic in $U(0, \epsilon)\backslash \Sigma_2$, such that
\begin{equation}\label{f-jump}f_+(\zeta)-f_-(\zeta) = |\zeta |^\alpha e^{\frac {2s} \zeta}~~\mbox{for}~\zeta \in (-\epsilon, 0).\end{equation}

\begin{figure}[t]
 \begin{center}
   \includegraphics[width=8cm]{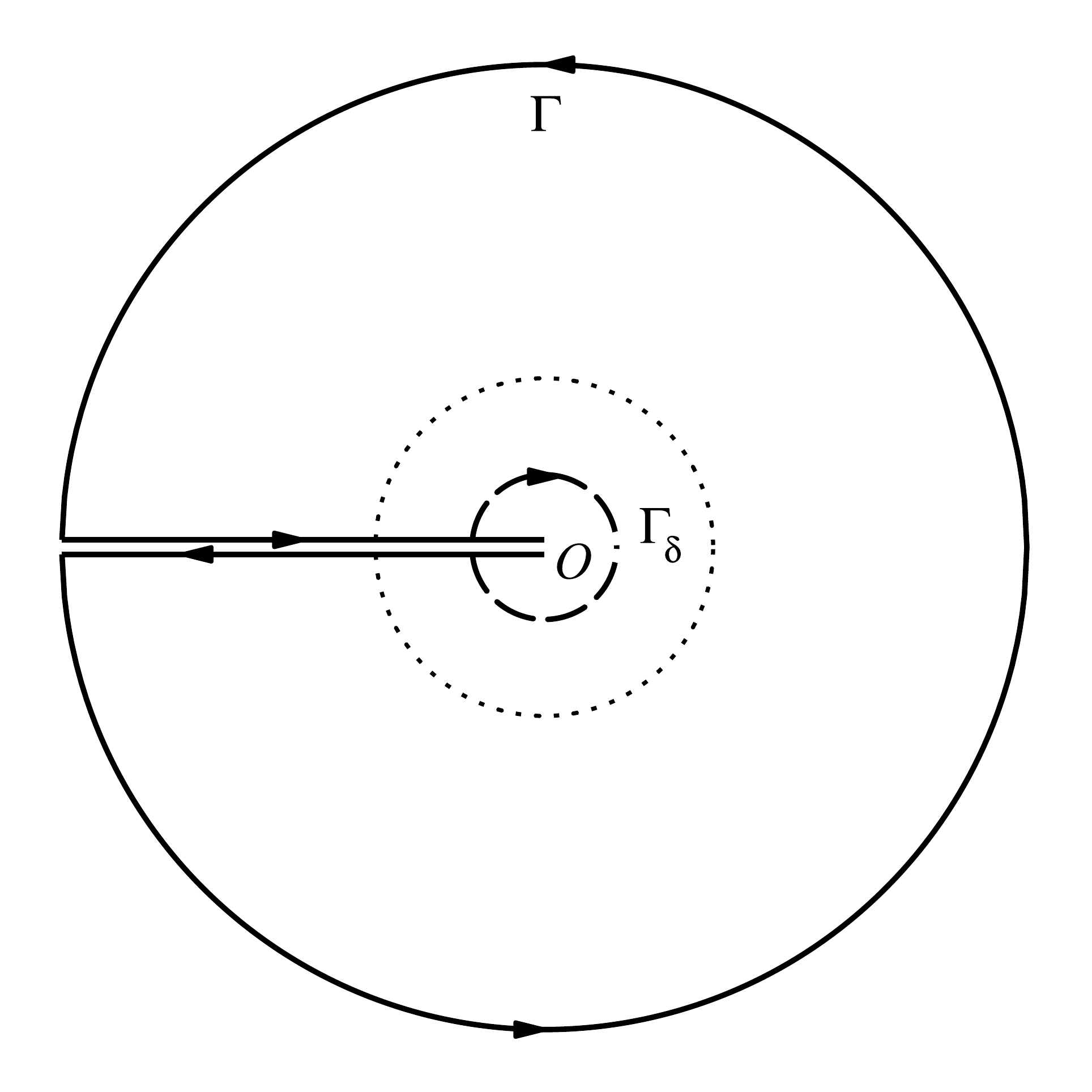} \end{center}
  \caption{\small{ The integration path $\Gamma$ in the complex $\tau$-plane: The solid bold path  consisting of the line segments along the upper and lower edge of $[-1, 0]$, and the circle $|\tau|=1$ joining them.
  $\Gamma_\delta$: The closed loop resulted from  replacing the line segments $[-\delta, 0]$ with the dashed bold circular part $|\tau|=\delta$. The dotted circle is $|\zeta|=\epsilon$, on which $\Phi(\zeta)$ and $F(\zeta)$ match.  }}
 \label{contour-for-integral}
    \end{figure}

Keeping in mind the matching condition \eqref{F-matching}, we chose
\begin{equation}\label{f-solution}f(\zeta) = - \frac 1 {4\pi\sin(\alpha\pi)} \int_\Gamma \frac {\tau^\alpha e^{2s/\tau}  d\tau} {\tau-\zeta},~~\zeta \in U(0,\epsilon)\backslash [-\epsilon, 0],\end{equation} where $\arg\tau\in [-\pi, \pi]$ in the integrand, the integration path $\Gamma$ consisting of the line segments along the upper and lower edge of $[-1, 0]$, and the circle $|\tau|=1$ joining them, as illustrated in Figure \ref{contour-for-integral}. We proceed to show that $f(\zeta)$ is what we are looking for.

First, we see that the jump condition \eqref{f-jump} is satisfied. Indeed, we may separate the path and rewrite
$$f(\zeta)=f_A(\zeta)-\frac 1 {4\pi\sin(\alpha\pi)} \left [ \int^0_{-\epsilon} \frac  { \left(|\tau|e^{i\pi}\right )^\alpha e^{2s/\tau} d\tau} {\tau-\zeta}- \int^0_{-\epsilon} \frac  { \left(|\tau|e^{-i\pi}\right )^\alpha e^{2s/\tau} d\tau} {\tau-\zeta}   \right ], $$ or, in a compact form
$$f(\zeta)=f_A(\zeta)+\frac 1 {2\pi i }   \int^0_{-\epsilon} \frac  { |\tau|^\alpha  e^{2s/\tau} d\tau} {\tau-\zeta}, $$
where $f_A(\zeta)$ denotes the integral on $\Gamma\backslash [-\epsilon, 0]$. Clearly, the  Cauchy integral  $f_A(\zeta)$ is analytic in $U(0,\epsilon)$. The jump condition \eqref{f-jump} follows accordingly from the Plemelj-Sokhotski  formula.

Next, we show that
\begin{equation}\label{f-approx}f(\zeta) =   \frac {\zeta^\alpha} {2i\sin(\alpha\pi)}  \left [1
+O\left ( \frac s \epsilon\right )+
O\left (\left (\frac \delta\epsilon\right )^{\alpha+1}\right )\right ],~~ |\zeta|=\epsilon,\end{equation} as
$s/\delta=O(1)$ and $\alpha>-1$, and the error term is uniform on the circle. To this aim, we deform the integration path $\Gamma$ in \eqref{f-solution} to a closed loop $\Gamma_\delta$ by replacing the line segments $[-\delta, 0]$ with the circular part $|\tau|=\delta$, where $\delta$ may depend on $s$, such that $0<\delta\ll \epsilon <1$; cf. the dashed circle in Figure \ref{contour-for-integral}. As a result one has
\begin{equation}\label{f-approx-path-deformed}f(\zeta) = f_\delta(\zeta)+ O\left (\epsilon^{-1}\delta^{\alpha+1}\right ),~~ |\zeta|=\epsilon,\end{equation}
where
$$f_\delta (\zeta) = - \frac 1 {4\pi\sin(\alpha\pi)} \int_{\Gamma_\delta} \frac {\tau^\alpha e^{2s/\tau}  d\tau} {\tau-\zeta},$$ and the integrals on the segment $\tau\in [-\delta, 0]$ and $|\tau|=\delta$  contribute to the error term in \eqref{f-approx-path-deformed}.  Indeed,
$$\left |\int ^0_{-\delta} \frac {\tau^\alpha e^{2s/\tau} d\tau } {\tau-\zeta}\right | \leq \frac 1 {\epsilon -\delta } \int^0_{-\delta} |\tau|^\alpha e^{-2s/|\tau|} d\tau\leq  \frac {\delta^{\alpha+1}} {\epsilon -\delta }e^{\frac {2s}\delta}=O\left (\epsilon^{-1}\delta^{\alpha+1}\right ), $$
so long as $\delta \ll \epsilon$ and $s/\delta=O(1)$. We obtain the same estimate for the circular part $|\tau|=\delta$. The formula \eqref{f-approx-path-deformed} is thus justified.

Using Cauchy's integral formula, for $\zeta\in \partial U(0, \epsilon)$, we have
\begin{equation}\label{f-delta-approx}f_\delta(\zeta)=\frac 1 {2i\sin(\alpha\pi)} \zeta^\alpha e^{2s/\zeta} = \frac {\zeta^\alpha} {2i\sin(\alpha\pi)}   \left ( 1+O\left ( \frac s \epsilon\right )\right),\end{equation}
where the branch is chosen such that $\arg \zeta\in (-\pi, \pi)$.
A combination of \eqref{f-approx-path-deformed} and \eqref{f-delta-approx}
gives  \eqref{f-approx}.

Noting that $e^{\frac s \zeta\sigma_3}=I+ O(s/\epsilon)$ for $|\zeta|=\epsilon$, substituting \eqref{f-approx} into \eqref{F-solution} and in view of \eqref{Phi-Bessel-origin-equal}, we see that the matching condition \eqref{F-matching} is well fulfilled. Here we have set $\delta=s$ and $\epsilon$ a small positive constant not depending   on $s$.
    Hence  $F(\zeta)$ defined in \eqref{F-solution} does  solve the  RH problem.

{\rem{\label{rem-alpha}  In deriving \eqref{f-approx}, we only need  $\alpha >-1$.
When $\alpha$ is a nonnegative integer, a logarithmic behavior may occur for the Bessel model problem $\Phi(\zeta)$ at the origin. Indeed, instead of  \eqref{Phi-Bessel-origin-equal}, now the behavior at $\zeta=0$ is
 $$\Phi(\zeta)=\hat{\Phi}(\zeta)\zeta^{\frac{\alpha\sigma_3}{2}}\left(
                                                          \begin{array}{cc}
                                                            1 &\frac{i}{\pi} (-1)^{\alpha+1} \ln  \left ( {\sqrt\zeta} /2\right )  \\
                                                            0 & 1 \\
                                                          \end{array}
                                                        \right) J,~~\arg\zeta\in (-\pi, \pi),~~\zeta\rightarrow 0,$$where $J$ is the same as in \eqref{Phi-Bessel-origin-equal}, and $\hat{\Phi}(\zeta)$  is an entire matrix function, as can be explicitly  determined by the ascending series of the modified Bessel functions, cf. \cite[(10.25.2), (10.31.1)]{nist}.
                                                        Still,  we can express the parametrix $F(\zeta)$ in the form of \eqref{F-solution}, with
   $$f(\zeta) =   \frac {(-1)^{\alpha+1}} {2\pi^2} \int_\Gamma \frac {\tau^\alpha e^{2s/\tau} \ln (\sqrt\tau /2) d\tau} {\tau-\zeta},~~\zeta \in U(0,\epsilon)\backslash [-\epsilon, 0],   $$
where $\Gamma$ is the same integration path employed in \eqref{f-solution}.   Following the steps  \eqref{F-solution}-\eqref{f-delta-approx},  it is readily verified that such a function $F(\zeta)$ solves the RHP \eqref{F-jump}-\eqref{F-origin} for   $\alpha=0,1,2,\cdots$. Slight modifications are needed here, for  example, \eqref{f-approx} now reads
$$ f(\zeta)= \frac {(-1)^\alpha} {\pi i} \zeta^\alpha \ln \left( \frac {\sqrt\zeta} 2\right ) \left (1+ O\left (\frac \delta \epsilon\right )+ O\left (\left (\frac \delta \epsilon\right )^{\alpha+1} \ln \delta \right )
\right )~~\mbox{as}~~|\zeta|=\epsilon,$$ and, accordingly, the matching condition \eqref{F-matching} now takes the form
$$F(\zeta)=\left [I+O(s)+O\left (s^{\alpha+1}\ln s\right )\right ]\Phi(\zeta)~~\mbox{for}~~|\zeta|=\epsilon,~~\mbox{as}~s\to 0^+.$$
}}

 {\rem{\label{rem-choosing-epsilon-delta}
We have options in choosing $\epsilon$ and $\delta$.  The previous estimation works as long as $s/\epsilon \ll 1$, $\delta/\epsilon \ll 1$, and $s/\delta=O(1)$, as $s\to 0^+$.  Here we  choose $\delta=s$, and $\epsilon$ small fixed. The choice makes sense since the corresponding Bessel functions, upon them the Bessel kernel is built, have no zeros in $U(0, \epsilon)$ for $\epsilon$ small enough. Then the approximation  of $\Psi(\zeta, s)$ in  $U(0, \epsilon)$ is in a sense of no significance, and one may focus on the Bessel-type approximation outside of the neighborhood $U(0, \epsilon)$.
 }}\vskip .3cm

Finally,  we  consider
\begin{equation}\label{R-0-def}R_0(\zeta)=\left\{\begin{array}{ll}
                                                   \Psi(\zeta,s)\Phi^{-1}(\zeta),   & |\zeta|> \epsilon,\\[.2cm]
                                                   \Psi(\zeta,s)F^{-1}(\zeta),       &|\zeta| <\epsilon.
                                                   \end{array}
\right.\end{equation}
The matrix function $R_0(\zeta)$ is analytic in $|\zeta|\neq \epsilon$, approaching $I$ at infinity,  and the jump on the circle is
\begin{equation}\label{R-0-jump}J_{R_0}(\zeta)=
                     I+ O(s )+O\left (s^{\alpha+1}\right ),\quad  |\zeta|= \epsilon;
\end{equation}see Figure \ref{contour-for-pro1} for the circular contour.
So, by an argument   similar to   Section \ref{sec:3.7}, we have
\begin{equation}\label{R-0-behavior}
 R_0(\zeta)=
 \left\{
 \begin{array}{ll}I+  O\left(s^\mu \right ),& s\rightarrow 0^+ ,~\mbox{uniform~for~bounded}~\zeta ,  \\[.2cm]
I+O\left (s^\mu \zeta^{-1}\right ),&\zeta\rightarrow\infty,~s\rightarrow 0^+,              \end{array}\right .
\end{equation}where $\mu=1$ for $\alpha >0$, and $\mu=\alpha+1$ for $-1<\alpha<0$.

This completes the nonlinear steepest descend analysis of $\Psi(\zeta,s)$ as $s\rightarrow 0^+$.

\subsection{Proof of Theorem \ref{thm-transition-to-Bessel}}

Now we have obtained approximations to    $\Psi(\zeta,s)$ as  $s\rightarrow 0^+$.  In the present subsection, we will reduce  $\Psi(\zeta,s)$ to the solution $\Phi(\zeta)$ of the Bessel model problem  for small $s$, and, as a by-product,   obtain the initial conditions for the nonlinear equations \eqref{equation-sets} and \eqref{r-eqn}, derived from the compatibility conditions of the Lax pair of  $\Psi(\zeta,s)$; cf. \eqref{psi-lax-1} and \eqref{psi-lax-2}.

It follows from (\ref{Phi-Bessel}), (\ref{R-0-def}) and (\ref{R-0-behavior}) that
\begin{equation}\label{Psi-in-Bessel}\Psi(\zeta,s)=\left [ I+O\left (s^\mu/\zeta \right)\right ]\Phi(\zeta),~~s\rightarrow 0^+,\end{equation}
for $ |\zeta|>\epsilon$, where $\mu=1$ for $\alpha\geq 0$ and $\mu=\alpha+1$ for $-1<\alpha<0$, $\epsilon$ is a small positive constant, and  $\Phi(\zeta)$ is the solution to the model Bessel problem, explicitly given in \eqref{Phi-Bessel}. Then a combination of \eqref{Phi-Bessel-infty} with \eqref{Psi-in-Bessel} gives
\begin{equation}\label{Psi-in-Bessel-approx}
\begin{array}{rcl}
\Psi(\zeta,s)e^{-\sqrt{\zeta}\sigma_3}M^{-1} \zeta^{\frac{1}{4}\sigma_3} &=&   I+\frac {4\alpha^2-1}{128\zeta}\left(
                                           \begin{array}{cc}
                                             4\alpha^2-9 & 16i \\
                                            \frac {i}{12} (4\alpha^2-9)(4\alpha^2-13) & 9-4\alpha^2 \\
                                           \end{array}
                                         \right) \\[.4cm]
                                   & &   +O\left (   {s^{\mu}}/ \zeta\right ) +O\left (\zeta^{-  3/2}\right)
                                                                                                              \end{array}
\end{equation}
as $\zeta\rightarrow\infty$, uniformly for $s\in(0,\varepsilon]$  with $\varepsilon$ small positive,  where  $M=(I+i\sigma_1)/{\sqrt{2}}$, $\mu=1$ for $\alpha\geq 0$ and $\mu=\alpha+1$ for $\alpha\in (-1, 0)$.

Let $s\rightarrow 0^+$, in view of \eqref{Psi-infty} and  \eqref{psi-c1-def}, we obtain the initial condition for the unknown functions $q(s)$, $r(s)$ and $t(s)$, as follows:
{\cor{\label{cor-initial-value}
The initial values    for the nonlinear equations for $q(s)$, $r(s)$ and $t(s)$ in  \eqref{equation-sets} and \eqref{r-eqn} can be determined, namely, \begin{equation}\label{initial-condition}
\left\{\begin{array}{l}
         q(0)=\frac 1 {128} {(4\alpha^2-1)(4\alpha^2-9)}\\[.2cm]
         r(0)=\frac 1 8\left (1- 4\alpha^2\right ) \\[.2cm]
         t(0)=\frac 1 {1536}(4\alpha^2-1)(4\alpha^2-9)(4\alpha^2-13).
       \end{array}\right.
 \end{equation}
}}

To complete the proof of Theorem \ref{thm-transition-to-Bessel}, we substitute  \eqref{Phi-Bessel} and \eqref{R-0-def} into \eqref{psi1-psi2-def}, and obtain
$$ \begin{array}{rcl}
  \begin{pmatrix}
     \psi_1(\zeta,s) \\
     \psi_2(\zeta,s) \\
  \end{pmatrix}  &= & R_0(\zeta) M_1\pi^{\frac {\sigma_3} 2}  e^{\frac {\pi i } 2 (\alpha-1)}
   \begin{pmatrix}
   I_\alpha ( \sqrt{|\zeta|}e^{-\frac {\pi i} 2})    \\
                                  \pi   \sqrt{|\zeta|}   I'_\alpha ( \sqrt{|\zeta|}e^{-\frac {\pi i} 2})
  \end{pmatrix}
   \\[.5cm]
     &= & R_0(\zeta)M_1\pi^{\frac {\sigma_3} 2}
     \begin{pmatrix}
     -i J_\alpha ( \sqrt{|\zeta|} )    \\
                                  \pi   \sqrt{|\zeta|}   J'_\alpha ( \sqrt{|\zeta|})
      \end{pmatrix},
   \end{array}  $$
where we have used the formula  $e^{\frac 1 2 \pi i \alpha}I_\alpha(z)=   J_\alpha (ze^{\frac 1 2\pi i})$ for $\arg z\in (-\pi, \pi/2]$; cf. \cite[(10.27.6)]{nist}, and an
  approximation for $R_0(\zeta)$ is provided  in \eqref{Psi-in-Bessel}.

We note that for an arbitrary matrix $\tilde M$ with  $\det \tilde M=1$, it holds
\begin{equation}\label{matrix-property} \tilde M^T \left(
                                                                                    \begin{array}{cc}
                                                                                      0 & -1 \\
                                                                                      1 & 0 \\
                                                                                    \end{array}
                                                                                  \right)\tilde M=\left(
                                                                                    \begin{array}{cc}
                                                                                      0 & -1 \\
                                                                                      1 & 0 \\
                                                                                    \end{array}
                                                                                  \right).\end{equation}
                   Here, as before, $\tilde M^T$ denotes the transpose of  $\tilde M$.
Substituting the above representation for $\psi_k$ into \eqref{kernel-approx-final} yields
\begin{equation}\label{kernel-Psi-Bessel-approx}
 \begin{array}{rcl}
   \displaystyle{ \frac {1}{4n} K_n\left (\frac {u}{4n} ,\frac {v}{4n}\right )}&=   & \displaystyle{ \frac{  \psi_1(-v,s)\psi_2(-u,s)   -  \psi_1(-u,s)\psi_2(-v,s)}{2\pi i(u-v)}+O\left (\frac 1 {n^2}\right)} \\[.4cm]
    &   = & \displaystyle{\mathbb{ J}_{\alpha}(u,v)+O\left (s^\mu\right )+O\left ( 1 /{n^2}\right),}
 \end{array}
  \end{equation}
  with $s=2nt=2nt_n\rightarrow 0^+$, $t$ is the parameter appeared in the weight \eqref{weight-of-the-paper}, where
  $\mathbb{ J}_{\alpha}$     is the Bessel kernel defined in \eqref{bessel-kernel}.
      The formula  in \eqref{kernel-Psi-Bessel-approx} holds  uniformly  for $u$ and $v$  in  compact subsets of  $(0,\infty)$ and uniformly for $s\in (0,\varepsilon]$.

Thus we complete the proof of Theorem \ref{thm-transition-to-Bessel}.

\section{Proof of Theorem \ref{thm-transition-to-Airy}: Transition to the Airy kernel  as $s\rightarrow+\infty$}\label{sec:6}
In this section, an  asymptotic analysis  of the model  RH problem for $\Psi(\zeta,s)$ is carried out  as the parameter
 $s\rightarrow \infty$. The results are  then applied to   the reduction of the $\Psi$-kernel in \eqref{kernel-approx-final} to the Airy kernel as $s\rightarrow\infty$. Attention will also be paid to  the large-$s$  asymptotics  of the equations in \eqref{r-eqn} and \eqref{equation-sets}.
 A similar discussion can be found in \cite{iko2009}.

\subsection{Nonlinear steepest descent analysis of  the model RH  problem as $s\rightarrow+\infty$}\label{sec:6.1}

\begin{figure}[h]
 \begin{center}
   \includegraphics[width=8cm]{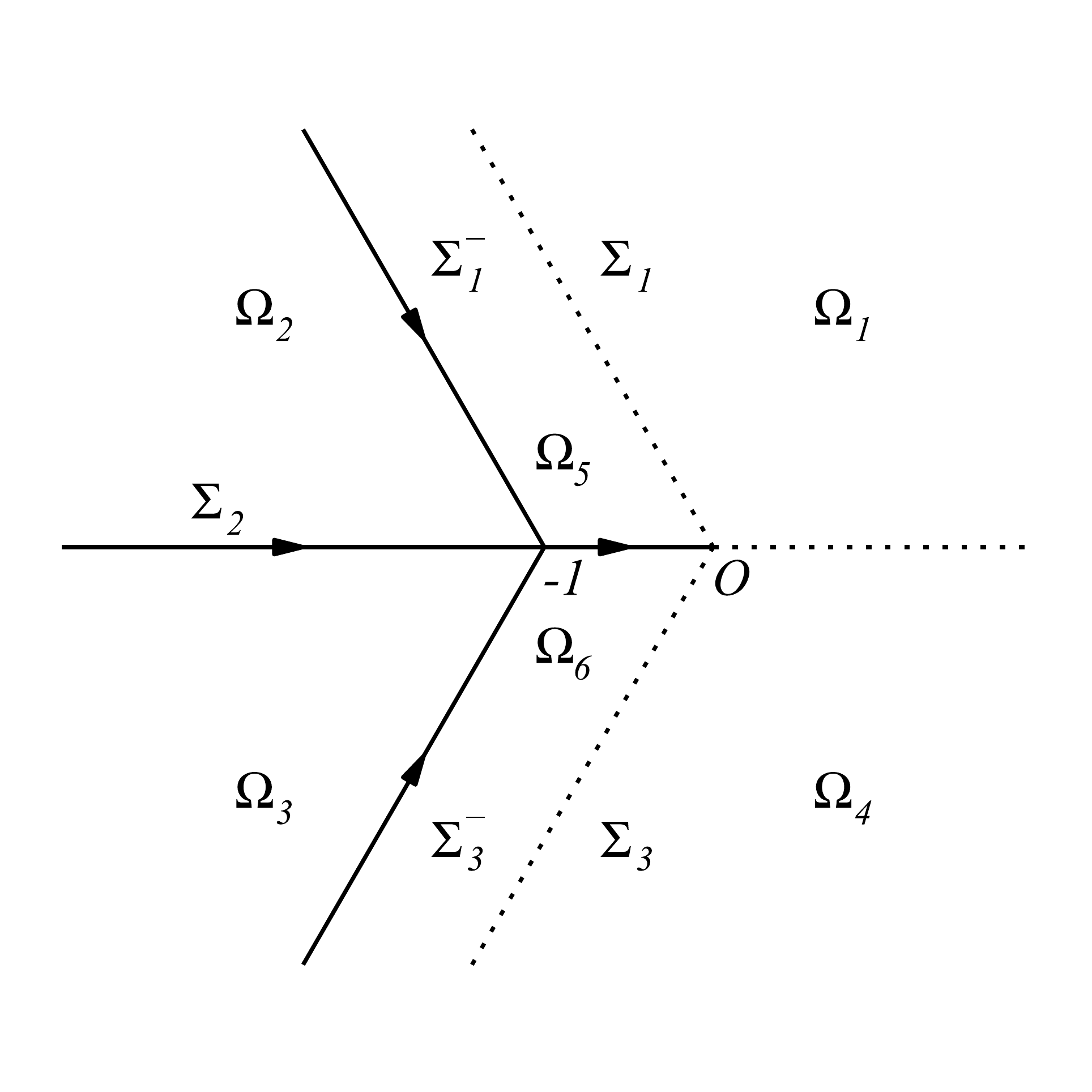} \end{center}
  \caption{\small{Contours for the $\Psi$ to Airy transition in the $\lambda$-plane, where $\Sigma_1^-$ and $\Sigma_3^-$ are obtained by respectively shifting $\Sigma_1$ and $\Sigma_3$ leftwards.}}
 \label{contour-for-pro2}
    \end{figure}

Taking a  normalization of  $\Psi(\zeta,s)$ (cf. \eqref{Psi-jump}-\eqref{Psi-origin}) at both the infinity and the origin, that is
\begin{equation}\label{transformation-Psi-to-U}
U(\lambda, s)=\left(
                \begin{array}{cc}
                  1 & 0 \\
                   \frac 32 i s^{\frac 13} & 1 \\
                \end{array}
              \right)
s^{\frac 16\sigma_3}\Psi(s^{2/3}\lambda, s)e^{-s^{1/3}\theta(\lambda)\sigma_3}, \quad \theta(\lambda)=(\lambda+1)^{3/2}/\lambda,
 \end{equation}
where $\arg(\lambda+1)\in (-\pi, \pi)$ and $\arg \lambda \in (-\pi, \pi)$,  we see that $U(\lambda, s)$ ($U(\lambda)$, for short) solves  the following RH problem:
\begin{description}
  \item(a)~~  $U(\lambda)$ is analytic in
  $\mathbb{C}\backslash\cup^3_{j=1}\Sigma_j$ (see  Figure \ref{contour-for-pro2});

  \item(b)~~  $U(\lambda)$  satisfies the jump conditions
  \begin{equation}\label{U-jump}
  U_+(\lambda)=U_-(\lambda)
  \left\{
  \begin{array}{ll}
     \left( \begin{array}{cc}
                                 1 & 0 \\
                                 e^{\alpha i\pi} e^{-2s^{1/3}\theta(\lambda)} & 1 \\
                               \end{array}
                             \right), &   \lambda \in \Sigma_1 ,  \\[.5cm]
     \left(
                               \begin{array}{cc}
                                 0 & 1 \\
                                 -1 & 0 \\
                                 \end{array}
                             \right),&  \lambda \in (-\infty,-1),  \\[.5cm]
      \left(
                               \begin{array}{cc}
                                 0 & e^{2s^{1/3}\theta(\lambda)} \\
                                 -e^{-2s^{1/3}\theta(\lambda)} & 0 \\
                                 \end{array}
                             \right), &   \lambda \in (-1,0) , \\[.5cm]
    \left( \begin{array}{cc}
                                 1 & 0 \\
                                 e^{-\alpha i\pi} e^{-2s^{1/3}\theta(\lambda)} & 1 \\
                               \end{array}
                             \right), &   \lambda \in \Sigma_3; \\
  \end{array}\right .
  \end{equation}

\item(c)~~  The asymptotic behavior of $U(\lambda)$  at infinity
  is
  \begin{equation}\label{U-infty}U(\lambda)=
    \left (I+O\left (\frac 1{\lambda}\right )\right)\lambda^{-\frac{1}{4}\sigma_3}\frac{I+i\sigma_1}{\sqrt{2}}
  ;\end{equation}
\item(d)~~The behavior of $U(\lambda)$  at the origin is, as $\lambda\to 0$,
\begin{equation}\label{U-origin}
U(\lambda)= O(1) \lambda^{\frac \alpha 2 \sigma_3} \left\{
\begin{array}{lr}
  I, & \lambda\in \Omega_1\cup\Omega_4, \\[.5cm]
  \left( \begin{array}{cc}
                                 1 & 0 \\
                                 -e^{\alpha i\pi} e^{-2s^{1/3}\theta(\lambda)} & 1 \\
                               \end{array}
                             \right), & \lambda\in \Omega_5, \\[.5cm]
   \left( \begin{array}{cc}
                                 1 & 0 \\
                                 e^{-\alpha i\pi} e^{-2s^{1/3}\theta(\lambda)} & 1 \\
                               \end{array}
                             \right), & \lambda\in \Omega_6;
\end{array}
\right .
\end{equation}see Figure \ref{contour-for-pro2} for the regions involved.
\end{description}

The second transformation $U\to X$ is to move  the jumps on $\Sigma_1$ and $\Sigma_3$ to   contours passing through  $\lambda=-1$, defined as
\begin{equation}\label{transformation-U-to-X}
X(\lambda)=X(\lambda, s)=\left\{\begin{array}{lr}
                    U(\lambda)\left( \begin{array}{cc}
                                 1 & 0 \\
                                 e^{\alpha i\pi} e^{-2s^{1/3}\theta(\lambda)} & 1 \\
                               \end{array}
                             \right), & \lambda\in \Omega_5 \\[.4cm]
                    U(\lambda)\left( \begin{array}{cc}
                                 1 & 0 \\
                                 -e^{-\alpha i\pi} e^{-2s^{1/3}\theta(\lambda)} & 1 \\
                               \end{array}
                             \right), & \lambda\in\Omega_6, \\[.4cm]
                   U(\lambda), & \mbox{otherwise}.
                  \end{array}
\right.
\end{equation}
Then  $X$ solves  the RH problem:
\begin{description}
  \item(a)~~  $X(\lambda)$ is analytic in
  $\mathbb{C}\backslash\Sigma_X$ ($\Sigma_X=(-\infty, -1)\cup (-1, 0)\cup \Sigma_1^-\cup \Sigma_3^-$; see Figure \ref{contour-for-pro2});

  \item(b)~~  $X(\lambda)$  satisfies the jump conditions
  \begin{equation}\label{X-jump}
  X_+(\lambda)=X_-(\lambda)
  \left\{\begin{array}{lr}
             \left( \begin{array}{cc}
                                 1 & 0 \\
                                 e^{\alpha i\pi} e^{-2s^{1/3}\theta(\lambda)} & 1 \\
                               \end{array}
                             \right), &  \lambda \in  \Sigma_1^-, \\[.4cm]
            \left(
                               \begin{array}{cc}
                                 0 & 1 \\
                                 -1 & 0 \\
                                 \end{array}
                             \right), &  \lambda \in (-\infty,-1), \\[.4cm]
                             \left( \begin{array}{cc}
                                 e^{\alpha i\pi} &  e^{2 s^{1/3}\theta(\lambda)} \\
                                0 &  e^{-\alpha i\pi} \\
                               \end{array}
                             \right),& \lambda \in (-1,0),\\[.4cm]
          \left( \begin{array}{cc}
                                 1 & 0 \\
                                 e^{-\alpha i\pi} e^{-2s^{1/3}\theta(\lambda)} & 1 \\
                               \end{array}
                             \right),&\lambda \in \Sigma_3^-;
                  \end{array}\right .
    \end{equation}

\item(c)~~  The asymptotic behavior of $X(\lambda)$  at infinity
  is kept
  \begin{equation}\label{X-infty}X(\lambda)=
   \left (I+O\left (\frac 1{\lambda}\right )\right)\lambda^{-\frac{1}{4}\sigma_3}\frac{I+i\sigma_1}{\sqrt{2}},~~\arg\lambda\in (-\pi, \pi),~\lambda\to\infty;\end{equation}
\item(d)~~The behavior of $X(\lambda)$  at the origin is
\begin{equation}\label{X-origin}
X(\lambda)= O(1)\lambda^{\frac \alpha 2\sigma_3}, ~~\arg\lambda\in (-\pi, \pi), ~ \lambda\to 0.
\end{equation}
\end{description}\vskip .5cm

Ignoring the   exponentially  small entries in the jumps  \eqref{X-jump}, we get an approximating   RH problem:
\begin{description}
  \item(a)~~  $\tilde{N}(\lambda)$ is analytic in
  $\mathbb{C}\backslash(-\infty,0)$ (see Figure \ref{contour-for-pro2});

  \item(b)~~  $\tilde{N}(\lambda)$  satisfies the jump conditions
   \begin{equation}\label{N-tilde-jump}
   \tilde{N}_+(\lambda)= \tilde{N}_-(\lambda)
   \left\{\begin{array}{lr}
             e^{\pi i\alpha \sigma_3},  &  \zeta \in(-1,0), \\[.2cm]
            \left(
                               \begin{array}{cc}
                                 0 & 1 \\
                                 -1 & 0 \\
                                 \end{array}
                             \right),  &   \zeta \in (-\infty, -1);
          \end{array}\right .
    \end{equation}
\item(c)~~  The asymptotic behavior of $\tilde{N}(\lambda)$  at infinity
  is
  \begin{equation}\label{N-tilde-infty}\tilde{N}(\lambda)=
   \left (I+O\left (\frac 1{\lambda}\right )\right)\lambda^{-\frac{1}{4}\sigma_3}\frac{I+i\sigma_1}{\sqrt{2}};\end{equation}
\item(d)~~  The asymptotic behavior of $\tilde{N}(\lambda)$  at $\lambda=0$
  is
  \begin{equation}\label{N-tilde-origin}
  \tilde{N}(\lambda)=O(1)\lambda^{\frac \alpha 2\sigma_3}.
\end{equation}
\end{description}

The RH problem for $\tilde{N}(\lambda)$ is an analogue of the global parametrix $N(z)$ treated in Section \ref{sec:3.4}, and, just like $N$, a   solution to the approximating  RH problem can be constructed explicitly as
(cf. \cite[(2.16)]{iko2009})
\begin{equation}\label{N-tilde-solution}\tilde{N}(\lambda)=\left(
                                                   \begin{array}{cc}
                                                     1 & 0 \\
                                                      i \alpha & 1 \\
                                                   \end{array}
                                                 \right)
(\lambda+1)^{-\frac 14\sigma_3}M\left(\frac{\sqrt{\lambda+1}+1}{\sqrt{\lambda+1}-1}\right)^{-\frac \alpha 2 \sigma_3},\end{equation}
where $M=\frac 1{\sqrt{2}}(I+i\sigma_1)$, the branches are chosen as $\arg(\lambda+1)\in (-\pi, \pi)$,  $\arg\lambda \in (-\pi, \pi)$, and such that the last factor in \eqref{N-tilde-solution} behaves
  $$\left(\frac{\sqrt{\lambda+1}+1}{\sqrt{\lambda+1}-1}\right)^{-\frac \alpha 2 \sigma_3}=\left (1+\frac {\alpha^2}{2\lambda}\right ) I -\frac \alpha{\sqrt\lambda} \sigma_3 +O\left (\lambda^{-3/2}\right )~~\mbox{for~large}~\lambda.$$

Comparing $X(\lambda)$ and $\tilde{N}(\lambda)$, the jumps of them have only an exponentially small difference, yet $\tilde N$ has an extra singularity at $\lambda=-1$. Hence, to approximate $X(\lambda)$,  a local parametrix has to be constructed in a neighborhood, say,
  $U(-1,r)$,  of $\lambda=-1$, where $r$ is a sufficiently small positive constant. The parametrix shares the same jumps \eqref{X-jump} with $X$ in the neighborhood, bounded at $\lambda=-1$, and matches with $\tilde N(\lambda)$ on $|\lambda+1|=r$.

It is readily verified that such a parametrix can be represented as follows
\begin{equation}\label{P1-representation}P_1(\lambda)=E_1(\lambda)\Phi_A\left (s^{\frac 29}f_1(\lambda) \right )e^{-s^{\frac 1 3}\theta(\lambda)\sigma_3}e^{\pm\frac12\alpha i\pi\sigma_3} \quad\mbox{for}~ \pm \Im \lambda>0,~\lambda\in U(-1,r),\end{equation}
where
\begin{equation}\label{P1-conformal-mapping}f_1(\lambda)=\left (-\frac 32\theta(\lambda)\right )^{\frac 23}, ~~\mbox{such~that}~f_1(\lambda)\sim \left (\frac 32\right )^{\frac 23}(\lambda+1)~~\mbox{for}~\lambda\sim -1
\end{equation} serves as a conformal mapping in $U(-1,r)=\{~\lambda\, |\, |\lambda+1|<r\}$ for sufficiently small $r$,
and  $\Phi_A$ is a solution to the Airy model RH problem in Section \ref{sec:3.5}, expressed explicitly as
\begin{equation}\label{Airy-model-solution}
  \Phi_A(\zeta)=M_A\left\{
                 \begin{array}{ll}
                  \left(
                     \begin{array}{cc}
                        \Ai(\zeta) & \Ai(\omega^2 \zeta) \\
                   \Ai'(\zeta)&\omega^2 \Ai'(\omega^2 \zeta) \\
                      \end{array}
                 \right)e^{-\frac{\pi i}{6}\sigma_3}, & \zeta\in \Omega_1 \\[.4cm]
                       \left(
                     \begin{array}{cc}
                        \Ai(\zeta) & \Ai(\omega^2 \zeta) \\
                      \Ai'(\zeta)&\omega^2 \Ai'(\omega^2 \zeta) \\
                      \end{array}
                  \right)e^{-\frac{\pi i}{6}\sigma_3}\left(
                                                       \begin{array}{cc}
                                                    1 & 0 \\
                                                         -1 & 1 \\
                                                       \end{array}
                                                      \right)
                  , & \zeta\in \Omega_2 \\[.4cm]
                            \left(
                     \begin{array}{cc}
                         \Ai(\zeta) & -\omega^2 \Ai(\omega \zeta) \\
                        \Ai'(\zeta)&- \Ai'(\omega \zeta) \\
                     \end{array}
                  \right)e^{-\frac{\pi i}{6}\sigma_3}\left(
                                                       \begin{array}{cc}
                                                        1 & 0 \\
                                                      1 & 1 \\
                                                       \end{array}
                                                    \right)
                  , & \zeta\in \Omega_3 \\[.4cm]
                        \left(
                     \begin{array}{cc}
                    \Ai(\zeta) & -\omega^2 \Ai(\omega \zeta) \\
                        \Ai'(\zeta)&- \Ai'(\omega \zeta) \\
                      \end{array}
                  \right)e^{-\frac{\pi i}{6}\sigma_3}, & \zeta\in \Omega_4;
                 \end{array}
          \right.
   \end{equation}cf.
  \cite[(7.9)]{dkmv2},
where  $\omega=e^\frac{2\pi i}{3}$, and the constant matrix $M_A=\sqrt{2\pi} e^{\frac 1 6\pi i}  \left(
                                                                               \begin{array}{cc}
                                                                                 1 & 0 \\
                                                                                 0 & -i \\
                                                                               \end{array}
                                                                             \right)$. Here, as an illustration, we may use Figure \ref{contour-for-model} to describe the regions $\Omega_1-\Omega_4$.

What remains is the determination of the factor
 $E_1(\lambda)$, so that  $E_1(\lambda)$ is
 analytic in $U(-1,r)$ and     makes   $P_1(\lambda) \approx \tilde N(\lambda)$ on $|\lambda+1|=r$.  We choose
\begin{equation}\label{P1-analytic-factor}E_1(\lambda)=\tilde{N}(\lambda) e^{\mp\frac12\alpha i\pi\sigma_3}\frac{I-i\sigma_1}{\sqrt{2}}\left (s^{\frac 29}f_1(\lambda) \right )^{\frac{1}{4}\sigma_3} \quad\mbox{for}~ \pm \Im \lambda>0,~\lambda\in U(-1,r).\end{equation}
Indeed, straightforward verification gives $\left(E_1\right )_+(\lambda)=\left (E_1\right)_-(\lambda)$
for $\lambda\in(-1-r, -1+r)$, and  $E_1(\lambda)$ is bounded at  $\lambda=-1$, which implies that $E_1(\lambda)$ is analytic in $U(-1,r)$ for $r$ sufficiently small. Moreover, it is readily clarify that, with $E_1(\lambda)$ and $P_1(\lambda)$ so defined,  we have the matching condition
\begin{equation}\label{P1-N-tilde-matching}P_1(\lambda)\tilde{N}(\lambda)^{-1}=I+O\left (s^{- 1/{3}}\right ),~~|\lambda+1|=r.\end{equation}

To complete  the Riemann-Hilbert analysis, we introduce the final  transformation $X\to R_1$ as
\begin{equation}\label{transformation-X-to-R1}R_1(\lambda)=\left\{\begin{array}{ll}
                                                   X(\lambda,s)\tilde{N}^{-1}(\lambda),   & |\lambda+1|> r,\\[.2cm]
                                                  X(\lambda,s)P_1^{-1}(\lambda),        &|\lambda+1| <r.
                                                   \end{array}
\right.\end{equation}
The matrix function $R_1(\lambda)$ is analytic in $\mathbb{C}\setminus \Sigma_{R_1}$, where $\Sigma_{R_1}$ consists of the parts
$\Sigma_k^-\setminus U(-1,r)$ for $k=1,3$, and the circular part   $\partial U(-1,r)$, along with the line segment $(-1+r, 0)$; see Figure \ref{contour-for-pro2}. $R_1(\lambda)$ is perfectly normalized at infinity and at $\lambda=-1$, is also  of $O(1)$ at $\lambda=0$;   cf. \eqref{X-origin} and \eqref{N-tilde-origin}. When $s \to \infty$, the jump for $R_1(\lambda)$ on $\Sigma_{R_1}$ has the following behavior:
\begin{equation}\label{R1-jump}J_{R_1}(\lambda)=\left\{\begin{array}{ll}
                                                        I+O(s^{- 1/{3}}), &  |\lambda+1|= r, \\
                                                        I+O\left (e^{-cs^{  1/{3}}}\right ), &  \Sigma_{R_1}\setminus\partial U(-1,r),
                                                      \end{array}
\right.
\end{equation}
where $c$ is a positive constant. For instance, on  $(-1+r, 0)$, a combination of  \eqref{X-jump}, \eqref{N-tilde-solution} and \eqref{transformation-X-to-R1} yields
$$J_{R_1}(\lambda)=I+|\lambda|^\alpha e^{s^{1/3} \theta(\lambda)}  O\left ( 1   \right ) =  I+|\lambda|^\alpha \exp \left ( - \frac { (\lambda+1)^{3/2} s^{1/3}}{|\lambda|} \right )  O\left ( 1   \right ),$$ which is of the form given in \eqref{R1-jump}, uniformly on $(-1+r, 0)$. Similarly, jumps on other contours  can be  estimated.
  So, by an argument as in  Section \ref{sec:3.7}, we have
\begin{equation}\label{R1-approximation}R_1(\lambda)=I+O(s^{-  1/{3}}),~~s\rightarrow \infty,\end{equation}
where the error term  is uniform in $\lambda$,  being kept  away from $\Sigma_{R_1}$.
Furthermore, for large $\lambda$, we have
\begin{equation}\label{R1-approximation-large-lambda}R_1(\lambda)=I+O\left (s^{-\frac 1{3}}\lambda^{-1}\right ),~~s\rightarrow \infty,~\lambda\to \infty.  \end{equation}

This complete the nonlinear steepest descend analysis of $\Psi(\zeta,s)$ as $s\rightarrow \infty$.

\subsection{Proof of Theorem \ref{thm-transition-to-Airy}}\label{sec:6.2}
Now we apply the above  asymptotic results for  $\Psi(\zeta,s)$ as $s\rightarrow \infty$, obtained by conducting the nonlinear steepest descent analysis, to
achieve  the transition  of $\Psi$-kernel  to the Airy kernel, and to extract  asymptotics of  the nonlinear equations in \eqref{equation-sets} and \eqref{r-eqn}, derived from  the compatibility conditions of the Lax pair \eqref{psi-lax-1}-\eqref{psi-lax-2} of  $\Psi(\zeta,s)$.

Tracing back the transformations $\Psi(\zeta, s) \to U(\lambda, s)\to X(\lambda, s)\to R_1(\lambda)$; cf. (\ref{transformation-Psi-to-U}), (\ref{transformation-U-to-X})  and  (\ref{transformation-X-to-R1}), it follows from the approximation (\ref{R1-approximation}) that
\begin{equation}\label{Psi-representation-in-N-tilde-outside}\Psi(s^{2/3}\lambda,s)=s^{-\frac16\sigma_3}\left(
                                                                           \begin{array}{cc}
                                                                             1 &0 \\
                                                                            -\frac 32i s^{ 1/3}  & 1 \\
                                                                           \end{array}
                                                                         \right)
\left (I+O\left (s^{-  1/{3}}/\lambda\right )\right )\tilde{N}(\lambda)e^{s^{  1/3}\theta(\lambda)\sigma_3} \end{equation}
 for  large $\lambda$ and  large  $s$. Here use has been made of the fact that $e^{-2s^{1/3} \theta(\lambda)}$ is exponentially small for $\lambda \in \Omega_5\cup \Omega_6$ as $\lambda\to \infty$.

Substituting the behavior    \eqref{N-tilde-infty} at infinity  for   $\tilde{N}(\lambda)$ into    \eqref{Psi-representation-in-N-tilde-outside} yields
\begin{equation}\label{Psi-representation-outside}\Psi(s^{2/3}\lambda,s) e^{-s^{ 1/3}\sqrt{\lambda}\sigma_3}=(s^{2/3}\lambda)^{-\frac{1}{4}\sigma_3}\left [I+O\left (s^{\frac 13}/\sqrt{\lambda}\right )\right ]\frac{I+i\sigma_1}{\sqrt{2}}
\end{equation}
for $\lambda\rightarrow\infty$ and  $s\rightarrow\infty$.

Similarly, for $ |\lambda+1|<r$,  a combination of    \eqref{transformation-Psi-to-U}, \eqref{transformation-U-to-X}  and \eqref{transformation-X-to-R1}) gives
\begin{equation}\label{Psi-representation-in-P1-inside}\Psi(s^{2/3}\lambda,s)=s^{-\frac16\sigma_3}\left(
                                                                           \begin{array}{cc}
                                                                             1 & 0 \\
                                                                             -\frac 32 is^{1/3} & 1 \\
                                                                           \end{array}
                                                                         \right)
R_1(\lambda) P_1(\lambda)e^{s^{  1/3}\theta(\lambda)\sigma_3}  \end{equation}
for   $\arg(\lambda+1)\in (\frac 23 \pi, \pi )\cup (- \pi,-\frac 23 \pi ) $, i.e., for $\lambda\in \Omega_2\cup\Omega_3$; cf. Figure \ref{contour-for-pro2}, such that $|\lambda+1|<r$, where
$R_1(\zeta)=
 I+O\left (s^{-  1/{3}}\right )$; see \eqref{R1-approximation}, and
$P_1(\lambda)$   is
constructed in terms of the Airy function; see (\ref{P1-representation}). Similar formulas are also true for  $\lambda$ in other sectors.
Thus    $\Psi(s^{2/3}\lambda,s)$   is  represented   by the solution $\Phi_A$ to the  model Airy RH problem as
\begin{equation}\label{Psi-representation-in-PhiA-inside}\Psi(s^{2/3}\lambda,s)=s^{-\frac16\sigma_3}\left(
                                                                           \begin{array}{cc}
                                                                             1 & 0 \\
                                                                            - \frac 32 is^{ 1/3} & 1 \\
                                                                           \end{array}
                                                                         \right)
R_1(\lambda)E_1(\lambda)\Phi_A\left (s^{\frac 29}f_1(\lambda) \right )e^{\pm\frac12\alpha i\pi\sigma_3}, \end{equation}respectively for $\pm\Im \lambda>0$, where $|\lambda+1|<r$.

Similar to the derivation leading to \eqref{kernel-Psi-Bessel-approx}, we obtain \eqref{Psi-Airy-approx-introduction}. In deriving \eqref{Psi-Airy-approx-introduction}, we need to approximate, case by case, the function $\Psi(s^{2/3}\lambda, s)$ for $\zeta\in \Omega_3$, and for $\zeta\in \Omega_6$; cf. Figure \ref{contour-for-pro2}. Also, we put in use   the technique of extracting \eqref{kernel-approx-final} from \eqref{kernel-approx-almost-final}. It is noted that  let $\lambda=-1+\frac u {cs^{2/9}}$, then the phase variable $s^{2/9}f_1(\lambda)$ can be expanded into a Maclaurin series in $u$ for bounded $u$ and large $s$, and that
 $s^{2/9}f_1(\lambda)=u\left [1+O\left ( u/s^{2/9}\right )\right ]$; cf. \eqref{P1-conformal-mapping}.   Repeated use has been  made of the formula \eqref{matrix-property} as well.

Now we turn to a brief discussion of the
  asymptotic properties  of  the nonlinear equation  \eqref{r-eqn}.
For $\Re \lambda>0$, \eqref{transformation-U-to-X}, \eqref{transformation-X-to-R1} and  \eqref{R-approx} imply that
\begin{equation}\label{U-N-tilde-relation}
  U(\lambda)=\left (I+O\left (\frac { s^{-1/3}} \lambda\right )\right )\tilde{N}(\lambda) \quad\mbox{as}~ \lambda\rightarrow\infty.
\end{equation}

On the one hand, substituting  \eqref{Psi-infty} and  \eqref{psi-c1-def}   into \eqref{transformation-Psi-to-U}, we approximate $U(\lambda)$ for large $\lambda$
\begin{equation}\label{U-evaluation-use-Psi}
  U(\lambda)M^{-1} \lambda^{\frac {\sigma_3}4} = I +\frac 1 \lambda \left(
                                  \begin{array}{cc}
                                    -\frac 32  r+\frac 98 s^{\frac 23}+q s^{-\frac 23} & i \left\{\frac 32 s^{\frac 13}-r s^{-\frac 13} \right \} \\
                                   i \left\{ \frac {27}{16} s -(\frac 94 r+\frac 38 )  s^{\frac 13}+3 qs^{-\frac 13}+\frac { t}s\right\} &  \frac 32  r-\frac 98 s^{\frac 23}-q s^{-\frac 23}\\
                                  \end{array}
                                \right)+O\left (\frac 1  {\lambda^{2}}\right ).
\end{equation}
On the other hand, we expand $\tilde{N}(\lambda)$ in \eqref{N-tilde-solution} for large $\lambda$ as
\begin{equation}\label{N-tilde-evaluation-expanding}
 \tilde{N}(\lambda)M^{-1} \lambda^{ \frac 14\sigma_3 } = I +\frac 1 \lambda \left(
                                  \begin{array}{cc}
                                   \frac {\alpha^2} 2-\frac 14& i\alpha \\
                                  i\left\{  \frac 1 2 \alpha^3- \alpha \right \}& -\frac {\alpha^2} 2+\frac 14\\
                                  \end{array}
                                \right)+O\left (\frac 1 {\lambda^2}\right ).
\end{equation}
Putting all  these together, we obtain
\begin{equation}\label{r-asymptotics-at-plus-infty}
r(s)=\frac 32  s^{\frac 23}-\alpha s^{\frac 13}+O(1)~~\mbox{as}~s\to +\infty.
\end{equation}

Now we turn to the last part of the proof, that is, the $\Psi$-kernel to Airy kernel transition,  as  $s\to \infty$. To begin, we calculate a quantity
$\alpha_n$, in the variable $z$ used in Section \ref{sec:3}: $z=\alpha_n$ corresponding to $\lambda=-1$ via the (not re-scaled)  conformal mapping $\zeta=n^2\left\{\phi\left (\frac z{4n}\right )\right\}^2\approx -4n  z$ in \eqref{conformal-mapping}, and the re-scaling $\zeta=s^{2/3} \lambda$   in \eqref{transformation-Psi-to-U}.
We see that $\alpha_n = \frac {s^{2/3}}{4n}$  makes $n^2\left\{\phi\left (\frac {\alpha_n }{4n}\right )\right\}^2 \sim -s^{2/3}$.    We note that $\alpha_n\to 0$ as $n\to \infty$, since $s=2nt$ and  $t$ is bounded from above.

It is worth  pointing out that $\alpha_n$ plays a role in a direct calculation of the equilibrium measure of the weight \eqref{weight-of-the-paper}, and serves as a so-called  Mhaskar-Rahmanov-Saff (MRS) number, or sometimes termed a soft-edge of the spectrum. So, the  edge behavior of statistic quantities, such as
the large-$n$  limit of the kernel $K_n(x,y)$, requires further investigation; cf. \cite{xzz2011} for the determination of the soft-edges in a similar case.

Fortunately,   the steepest descent analysis conducted  earlier    in the present section has constructed an Airy type asymptotic approximation of $\Psi(s^{2/3}\lambda, s)$; see \eqref{Psi-representation-in-P1-inside} and \eqref{P1-representation}. The approximation is given in a normal-sized neighborhood of $\lambda=-1$, and, equivalently, a shrinking neighborhood of $z=\alpha_n$ of size $O(\alpha_n)$.

Substituting \eqref{Airy-model-solution}  and \eqref{Psi-representation-in-PhiA-inside} into the $\Psi$-kernel in \eqref{kernel-approx-final}, similar to the derivation leading to \eqref{kernel-Psi-Bessel-approx},
we have
\begin{equation}\label{kernel-Psi-Airy-approx}
\lim_{n\rightarrow \infty}\frac {\alpha_n}{cs^{2/9}} K_n\left (\alpha_n-\frac{\alpha_n}{cs^{2/9}}u ,\alpha_n-\frac{\alpha_n}{cs^{2/9}}v; t\right )=\frac{\Ai(u)\Ai'(v)-\Ai(v)\Ai'(u)}{u-v}
 \end{equation}
as $s=2nt \rightarrow +\infty$, where   $\alpha_n=s^{ 2/3}/(4n)$, $c=(\frac 32)^{ 2/3}$, and the limit is uniformly taken   for bounded $u,v\in \mathbb{R}$. This is exactly \eqref{kernel-Airy-approx-introduction}.

Thus we complete the proof of Theorem \ref{thm-transition-to-Airy}.

\section*{Acknowledgements}
The authors are very grateful to the anonymous reviewers  for their helpful and constructive comments. A reduction of the third-order equation to the PIII equation is completed in this version,   owing much to their suggestions.

 The work of Shuai-Xia Xu  was supported in part by the National
Natural Science Foundation of China under grant number
11201493, GuangDong Natural Science Foundation under grant number S2012040007824, Postdoctoral Science Foundation of China under Grant No.2012M521638, and the Fundamental Research Funds for the Central Universities under grand number 13lgpy41.
           Dan Dai was partially supported by a grant from City University of Hong Kong (Project No. 7004065) and grants from the Research Grants Council of the Hong Kong Special Administrative Region, China (Project No. CityU 100910, CityU 101411).
 Yu-Qiu Zhao  was supported in part by the National
Natural Science Foundation of China under grant number
10871212.

\end{document}